\documentclass{aa}  
\usepackage{natbib}
\usepackage{graphicx}
\usepackage{txfonts}
\usepackage{calc}
\usepackage{color}
\begin{document}
   \title{Brightness, distribution, and evolution of sunspot umbral dots}

   \author{T. L. Riethm\"uller
          \and
          S. K. Solanki
          \and
          V. Zakharov
          \and
          A. Gandorfer
          }

   \institute{Max-Planck-Institut f\"ur Sonnensystemforschung (MPS),
              Max-Planck-Str. 2, 37191 Katlenburg-Lindau, Germany\\
              \email{[riethmueller;solanki;zakharov;gandorfer]@mps.mpg.de}
             }

   \date{Received 29 July 2008 / Accepted 15 October 2008}

 
  \abstract
   {
   Umbral Dots (UDs) are thought to be manifestations of magnetoconvection in sunspot umbrae. Recent
   advances in their theoretical description point to the need for a thorough study of their properties
   and evolution based on data with the highest currently achievable resolution.
   }
   {
   Our UD analysis aims to provide parameters such as lifetimes, diameters, horizontal velocities,
   and peak intensities, as well as the evolution of selected parameters.
   }
   {
   We present a 106-minute TiO (705.7~nm) time series of high spatial and temporal resolution that
   contains thousands of UDs in the umbra of a mature sunspot in the active region NOAA~10667
   at $\mu$~=~0.95. The data were acquired with the 1-m Swedish Solar Telescope (SST) on La Palma.
   With the help of a multilevel tracking (MLT) algorithm the sizes, brightnesses, and trajectories
   of 12836 umbral dots were found and extensively analyzed. The MLT allows UDs with very low contrast
   to be reliably identified.
   }
   {
   Inside the umbra we determine a UD filling factor of 11\,\%. The histogram of UD lifetimes is
   monotonic, i.e. a UD does not have a typical lifetime. Three quarters of the UDs lived for less than 150~s
   and showed no or little motion. The histogram of the UD diameters exhibits a maximum at 225~km,
   i.e. most of the UDs are spatially resolved. UDs display a typical horizontal velocity of
   420~m\,s$^{-1}$ and a typical peak intensity of 51\,\% of the mean intensity of the quiet
   photosphere, making them on average 20\,\% brighter than the local umbral background. Almost all
   mobile UDs (large birth-death distance) were born close to the umbra-penumbra boundary, move towards
   the umbral center, and are brighter than average. Notably bright and mobile UDs were also observed along a
   prominent UD chain, both ends of which are located at the umbra-penumbra boundary. Their motion started
   primarily at either of the ends of the chain, continued along the chain, and ended near the chain's
   center. We observed the splitting and merging of UDs and the temporal succession of both. For the
   first time the evolution of brightness, size, and horizontal speed of a typical UD could be determined
   in a statistically significant way. Considerable differences between the evolution of central and
   peripheral UDs are found, which point to a difference in origin.
   }
   {}


   \maketitle

\section{Introduction}

   The investigation of the complex fine structure of umbrae and penumbrae is crucial
   to understanding the subsurface energy transport in sunspots. The energy transport from the
   solar interior to the solar surface outside magnetic features is mainly determined by convection,
   visible as granulation in images of the quiet photosphere. The strong and nearly vertical umbral
   magnetic field suppresses normal overturning convection inside the umbra. However, it is believed
   that some form of residual magnetoconvection is responsible for much of the remaining energy
   transport and manifests itself in the form of fine structures, such as light bridges (LBs) or
   umbral dots (UDs). In the present paper we consider UDs, which contribute up to 37\,\%
   of the radiative umbral flux according to \citet{Adjabshirzadeh1983}.
   Different models have been proposed to explain the umbral dots. \citet{Choudhury1986}
   postulated that UDs are thin columns of field-free hot gas between the cluster of small magnetic
   flux tubes that form the subsurface structure of a sunspot according to \citet{Parker1979}.
   According to this model, a UD is formed when an upwelling brings hot material into the photosphere.
   An alternative model has been proposed by \citet{Weiss1990} who consider UDs to be spatially
   modulated oscillations in a strong magnetic field.

   A more recent, promising approach is presented by \citet{Schuessler2006}, who used numerical simulations
   of three-dimensional radiative magnetoconvection to improve the physical understanding of the umbral
   fine structure. The simulations exhibit the emergence of small-scale upflow plumes that start off
   like oscillatory convection columns below the solar surface but turn into narrow overturning cells
   driven by the strong radiative cooling around optical depth unity. Most of those UDs show a central
   dark lane. The presence of dark lanes in penumbral and umbral fine structures has already
   been observed several times, cf. \citet{Scharmer2002,Langhans2007,Scharmer2007}. The verification of
   the predicted dark lanes in large UDs by \citet{Bharti2007} and \citet{Rimmele2008},
   as well as the verification of the predicted photospheric stratification of bright peripheral UDs
   by \citet{Riethmueller2008}, support the Sch\"ussler \& V\"ogler model of UDs. There
   is now a need to learn more about this phenomenon, with a statistically robust analysis of UD
   properties and evolution being a promising means of achieving this aim.

   The most detailed analyses of UDs are more than 10 years old \citep{Sobotka1997a,Sobotka1997b}
   and are based on data observed with the 50-cm SVST (Swedish Vacuum Solar Telescope), cf. the
   recent reviews of umbral fine structures by \citet{Solanki2003,Thomas2004} and \citet{Sobotka2006}.
   The more recent papers of \citet{Tritschler2002,Hartkorn2003} and \citet{Sobotka2005} have
   concentrated on individual properties and lack, e.g., the determination of UD trajectories.
   Furthermore, the possibility of improving the spatial resolution with the help of modern image
   reconstruction algorithms is only used by \citet{Tritschler2002}. The present paper aims to
   overcome these shortcomings, by employing data from the 1-m SST (Swedish Solar Telescope) equipped
   with an adaptive optics system, by restoring the data employing MFBD (multi-frame blind
   deconvolution), and determining the evolution of UD parameters whenever possible.


\section{Observations and data reduction}

   The data employed here were acquired on September 7, 2004 with the Swedish Solar Telescope
   at the Observatorio del Roque de los Muchachos on La Palma, Spain. Technical details of
   the SST are described by \citet{Scharmer2003a}. Wavefront aberrations caused by the telescope
   and by the turbulent atmosphere of the Earth were partially corrected by the adaptive optics system,
   explained in \citet{Scharmer2003b}. The science camera was a Kodak Megaplus CCD with a pixel size
   of 9 $\mu$m and a plate scale of 0.041$^{\prime\prime}$ (30~km on the Sun) per pixel. The camera was
   equipped with an interference filter at the wavelength of the 705.7~nm of the titanium
   oxide band head, the FWHM of this filter was 0.71~nm. The theoretical diffraction limit
   of the telescope at the TiO wavelength is 0.18$^{\prime\prime}$ (130km). Due to the high sensitivity
   to umbral temperatures of TiO lines, the TiO band head is a good diagnostic wavelength range for imaging
   umbral features \citep{Berdyugina2003}. A wavelength in the red was chosen also in order to ensure a more
   homogeneous time series due to the more benign seeing at these wavelengths. Acquisition lasted from
   08:27~UT to 10:17~UT, i.e. a total of 110~min. The images were obtained in a frame selection mode that
   saved only the 8 best images of a 20-second-interval. The exposure time was 10~ms. The telescope pointed
   to the sunspot of the active region NOAA~10667 at cos~$\theta$~=~0.95, i.e. relatively close to the solar
   disk center ($\theta$ is the heliocentric angle).

   The data were dark current and flat field corrected, reconstructed via the MFBD technique
   \citep{Loefdahl2003}, derotated, destretched \citep{November1988}, and subsonic filtered
   with a cut-off phase velocity of 5~km\,s$^{-1}$ \citep{Title1989}. The obtained
   time series consists of 310 images with a spatial resolution in the range of
   $\sim$0.18$^{\prime\prime}$-0.25$^{\prime\prime}$, as we estimated from radially averaged power spectra.
   The field of view (FOV) is 37$^{\prime\prime}$~$\times$~59$^{\prime\prime}$ and contains the entire
   considered sunspot whose umbra is divided into two parts by a light bridge. UDs in both
   parts of the umbra are analyzed in the next section.

\section{Data analysis}

\begin{figure}
   \centering
   \includegraphics[width=0.5\linewidth-1mm]{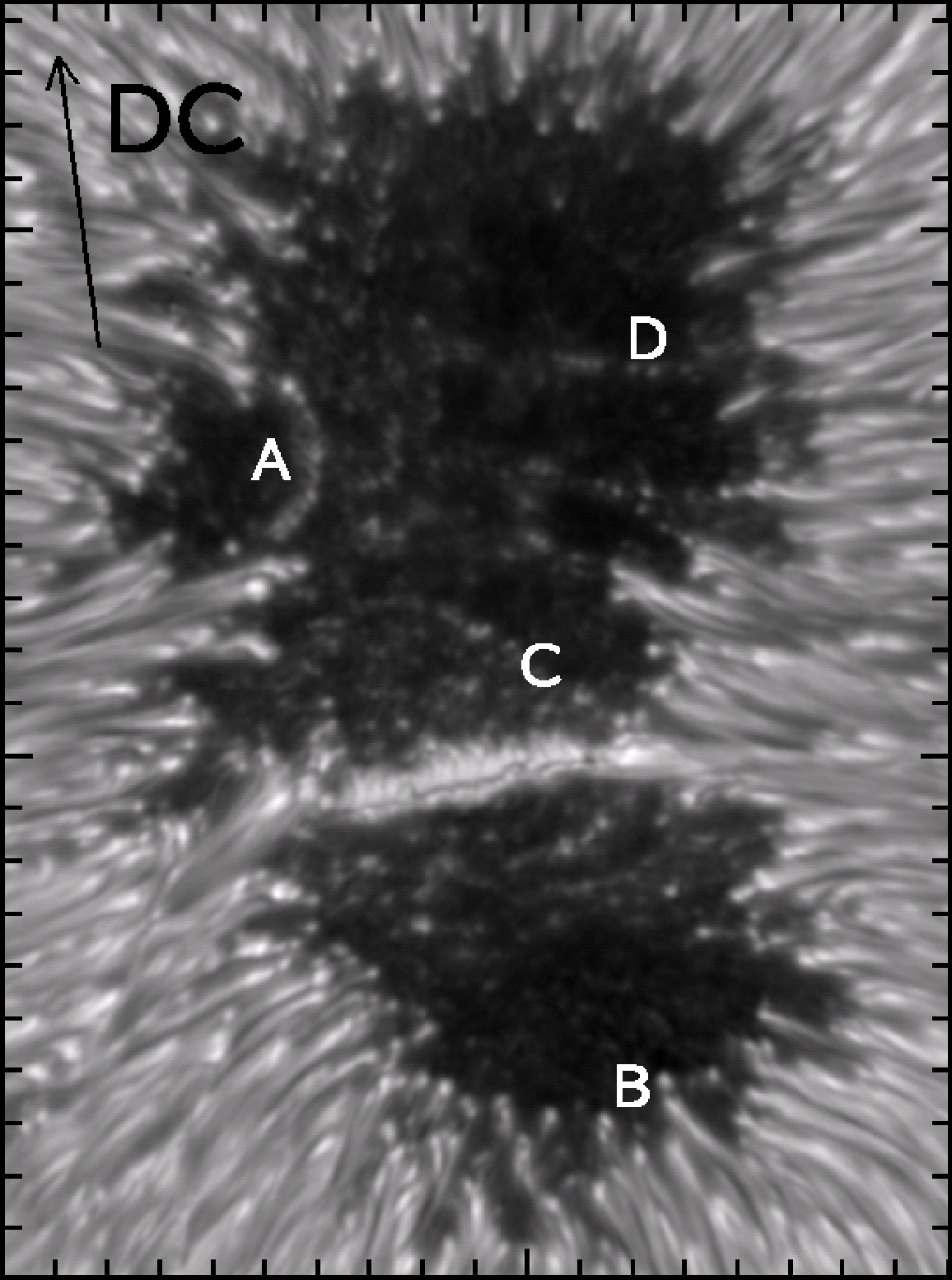}
   \includegraphics[width=0.5\linewidth-1mm]{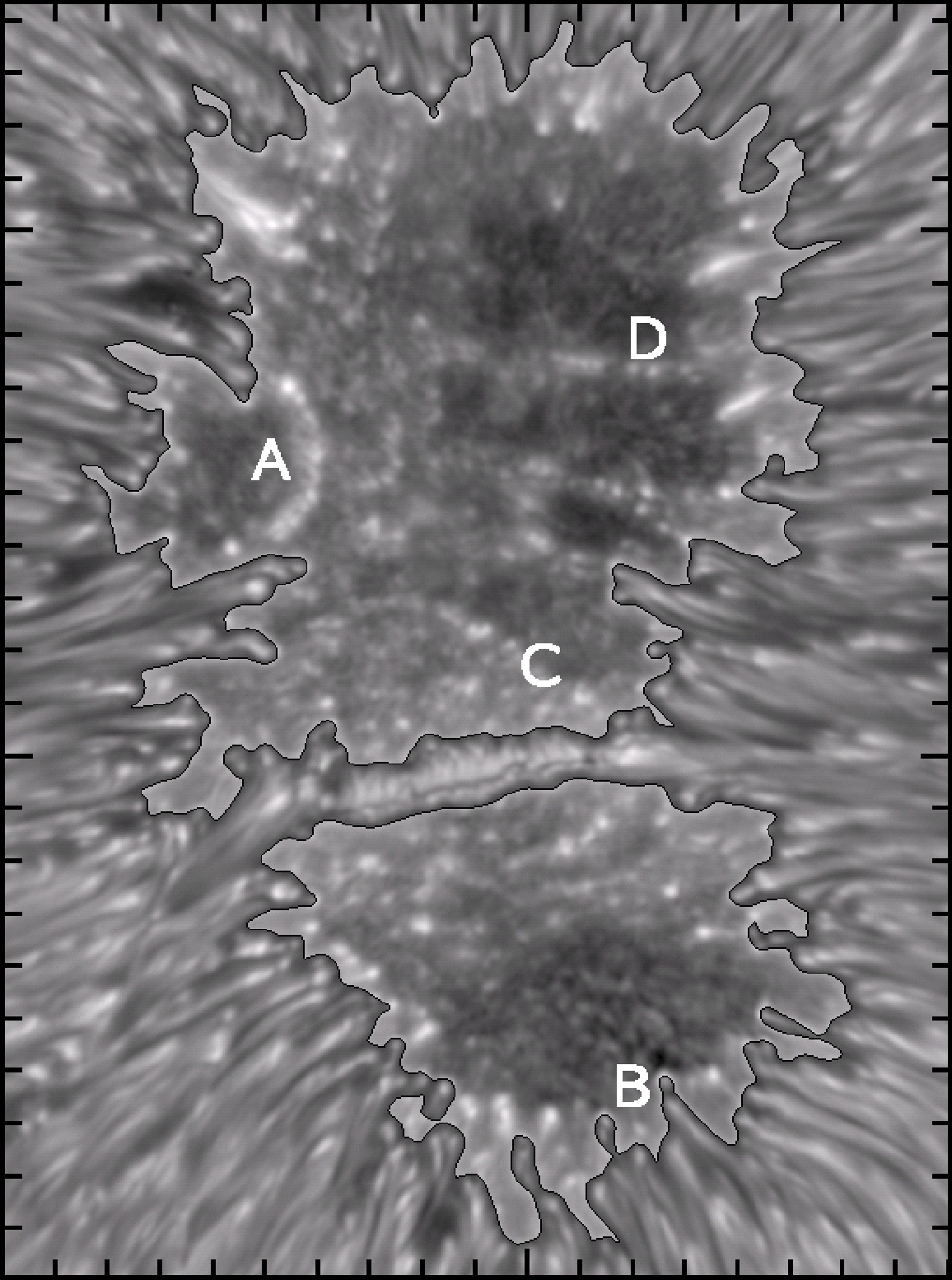}
   \caption{Frame taken at 09:53:20 UT: Best quality image plotted at normal contrast (left panel)
   and with increased umbral brightness (right panel). The black contour lines outline the umbra (see text
   for details). The direction to solar disk center (DC) is indicated by the arrow. Locations of
   umbral dots (UDs) that are discussed further in the text are marked by the letters A-D.
   Minor tick marks are given in Mm.}
   \label{FigBestImage}
   \end{figure}

   The detailed analysis of 310 images requires an automated algorithm for the identification of
   the thousands of umbral dots they contain. A specific algorithmic challenge is the fact
   that UDs as well as the local umbral background between them cover a broad range of intensities.
   At a normal contrast (left panel of Fig.~\ref{FigBestImage}) UDs are mainly visible near the
   penumbra. By displaying the square root of the umbral brightness instead of the
   brightness itself numerous UDs within the dark umbral background become visible as well
   (right panel).

   Our automated UD analysis starts with isolating the umbra, which is done by thresholding a lowpass
   filtered image (averaging a squared environment of 11~$\times$~11 pixels) at 35\,\% of the mean
   intensity of the quiet photosphere ($I_{ph}$). From the resulting set of contours we select only
   those longer than 3~Mm in order to avoid larger UDs from being connected to the umbral boundary.
   In the particular case of the studied penumbra the results are almost identical to identifying
   the two longest contours with the umbral boundary. The thus obtained umbral boundary is visible
   in the right panel of Fig.~\ref{FigBestImage} as the black contour line. This method for isolating
   the umbral boundary automatically ensures that a local brightening at the end of a penumbral fibril
   is only considered to be a UD if it is isolated from the penumbra in the sense that the intensity
   between the UD and the penumbral fibril falls below the applied threshold of 0.35~$I_{ph}$.

\begin{figure}
   \centering
   \includegraphics[width=\linewidth-1mm]{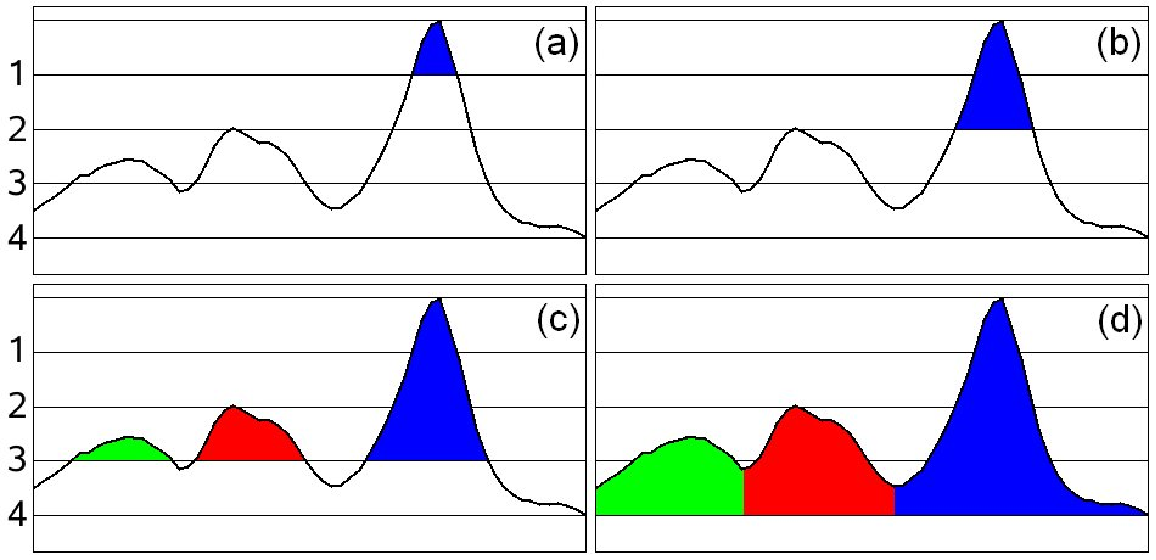}
   \caption{Illustration of multilevel tracking algorithm with 4 levels applied to a typical intensity profile.}
   \label{FigMLT}
   \end{figure}

   In the next step the UDs in each of the 310 images (recorded at a cadence of 20.57~s) are
   detected. For this purpose several algorithms were tested, e.g. a method where, starting
   from the UD center, 8 equally distributed rays are followed until they reach the UD boundary which
   is defined as the position where the intensity drops below 50\,\% of the maximum intensity above the
   local umbral background. The resulting 8 boundary points lead to a polygon that is a good approximation
   of the UD boundary. The method does not work properly for UDs with a partly concave boundary and it
   cannot easily separate UDs that are close to each other. Finally, the multilevel tracking
   algorithm of \citet{Bovelet2001}, which provided the best results in detecting the UD boundaries,
   was chosen. First the MLT algorithm determines the global extrema of the umbral intensities and
   subdivides this range into equidistant levels. Bovelet and Wiehr used MLT to distinguish between
   granules and intergranular lanes of the quiet Sun and found that three MLT levels are sufficient
   for their purpose. Since umbral dots cover a broad range of intensities we have to use a noticeably
   higher number of levels. We normalized our best quality image to $I_{ph}$ and found an umbral
   intensity range from 0.36-0.96~$I_{ph}$ (Note: This range is only valid for the best quality image,
   other images may reach lower or higher umbral intensities). We found that 25 MLT levels is the
   optimal compromise between detecting as many UDs as possible and avoiding the misinterpretation
   of noise as UDs. Whereas small umbral dots, obtained with this choice of levels, have a
   typical contrast relative to the local background of about 0.05~$I_{ph}$, the noise level
   is about 0.005~$I_{ph}$ (see Fig.~\ref{FigMLT} for a typical intensity profile).
   Starting with the highest intensity level all pixels are found whose intensity exceeds this level.
   This leads to several bounded two-dimensional structures, that are tagged in a unique way, which
   is indicated by different colors in the one-dimensional illustration given in Fig.~\ref{FigMLT}.
   The obtained closed structures are extended pixel by pixel as long as the intensity is greater
   than the next lower level. After that the algorithm searches through the whole umbra again to find
   all pixels whose intensity is greater than the next lower level, which may lead to some newly
   detected closed structures. This procedure is repeated until the minimal intensity level is reached.
   At the end every umbral pixel belongs to exactly one closed structure. The mode of operation of the
   MLT algorithm is illustrated for one dimension in Fig.~\ref{FigMLT} for the case of 4 levels.

\begin{figure}
   \centering
   \includegraphics[width=\linewidth-1mm]{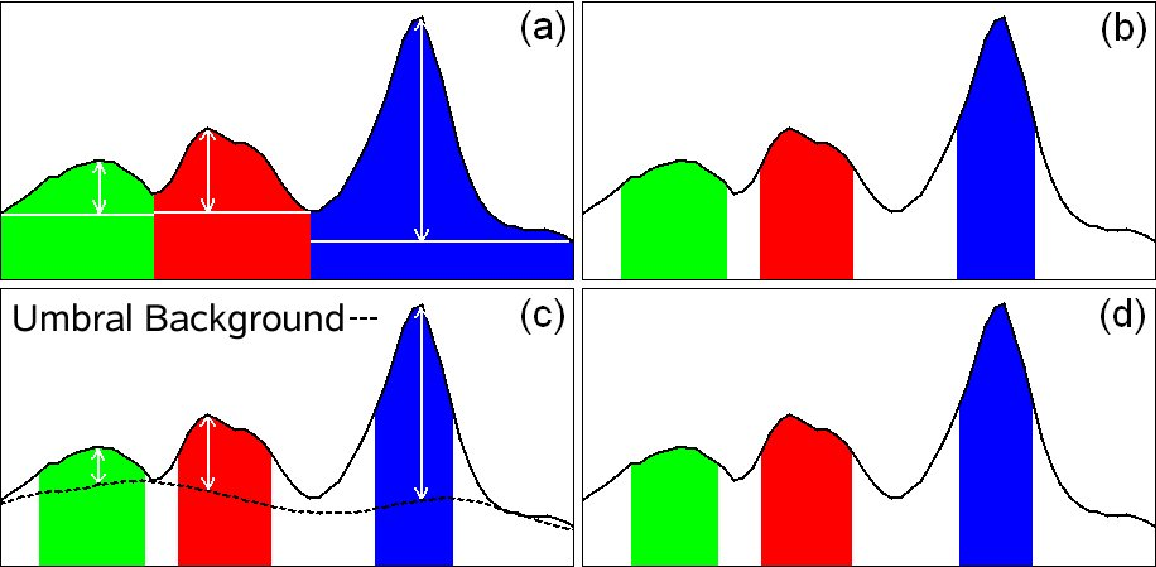}
   \caption{Determination of the boundary of umbral dots.}
   \label{FigCuttingPixels}
   \end{figure}

   The minimal ($I_{Min}$) and maximal ($I_{Peak}$) intensity of each closed structure is determined
   and all pixels that have an intensity lower than 50\,\% of this min-max range (white arrows in
   Fig.~\ref{FigCuttingPixels}a) are cut. This leads to a first estimate of the UD boundaries
   (see Fig.~\ref{FigCuttingPixels}b) that are used to determine the local umbral background intensities
   ($I_{bg}$), i.e. the intensities that would be observed in the absence of all UDs. We applied
   the method used by \citet{Sobotka2005} that approximates the local umbral background by a 2D surface
   fitted to the grid of local intensity minima, using the method of thin-plate splines
   \citep{Barrodale1993}. Since the local umbral background intensities are known now
   (dashed line in Fig.~\ref{FigCuttingPixels}c), we determine the exact UD boundaries by cutting
   all pixels lower than 50\,\% of the maximum intensity above the local umbral background (see white arrows in
   Fig.~\ref{FigCuttingPixels}c). Fig.~\ref{FigCuttingPixels}d illustrates that the resulting UD
   boundaries are similar to our first estimate in Fig.~\ref{FigCuttingPixels}b so that we don't need
   further iterations (this was the case with most identified UDs). Employing this procedure we found,
   on average, 323 UDs per image. Sometimes, our algorithm recognizes strong elongated bright structures
   as UDs that we would not consider as a UD by visual inspection of the images. Via a spot-check on
   selected images we estimate that the number of misidentifications is lower than 1\,\%. Due
   to the large number of detected UDs we are not able to remove the misidentifications by hand and
   accepted them as noise.

   Since a UD is an extended structure we determine the coordinates of the brightest pixel (peak intensity)
   and save them as the UD's position. This method is applicable because the noise was sufficiently reduced
   by our subsonic filter, as demonstrated by a typical intensity profile in Fig.~\ref{FigMLT}. We also
   determine the UD's diameter, defined as the diameter of a circle of area equal to that within the boundary
   of the UD. After the positions of all UDs of every image are known, the motions and trajectories of UDs
   are determined. The continuation of a trajectory in the image at the next (previous) time step is determined
   by finding the UD that is closest to the UD's current position. If no UD can be found within a 5~pixel
   neighborhood (theoretical diffraction limit) of the current UD then the tracking stops. In a loop over all images every UD is tracked
   backward in time until its birth and forward in time until its death. The tracking must be
   tolerant to the occasional image with lower image quality in which the UD may not be correctly identified
   (specially the smaller and fainter ones). In practice we allow for a gap of up to two images. If a UD is
   present at nearly the same location on both sides of the gap, the tracking is continued. In this
   manner we found 12836 UD trajectories that are weakly lowpass filtered in space (averaging the
   positions of the same UD in 15 consecutive images) in order to further reduce seeing-induced noise.
   From now, we call such a smoothed trajectory simply $trajectory$. We note that 5949 of the
   12836 UDs are only identified in a single image. We decided not to ignore them, because these bright dots in
   the umbra are detected rather well, even if only for a very short time. We assigned a zero trajectory length
   and a lifetime of 20.57~s to these 5949 UDs.

\section{Results}

   \subsection{Qualitative results}

   A first impression of the temporal evolution of the smallest umbral structures is reached by making
   a movie of the reconstructed time series of images. Some interesting phenomena are found by the
   visual inspection of this movie and are explained briefly below.

   The sunspot has two umbrae, a smaller and a roughly twice larger one separated by a light bridge (LB).
   The LB contains a clearly visible dark lane in agreement with the observations of, e.g.,
   \citet{Berger2003}. The lane is closer to the limbward edge of the LB. Possibly this is because the
   sunspot was observed at a heliocentric angle of $\theta~=~\mathrm{18^\circ}$, so that projection
   effects may cause the observed asymmetry \citep[e.g.][]{Lites2004}. However, the LB also displays
   another major asymmetry: the movie exhibits many UDs that are born within the LB and move into the
   larger umbra, i.e. towards the solar disk center (arrow in Fig.~\ref{FigBestImage}), while almost
   no UD leaves the LB into the smaller umbra, i.e. towards the solar limb (antiparallel to the arrow).

\begin{figure}
   \centering
   \includegraphics[width=\linewidth]{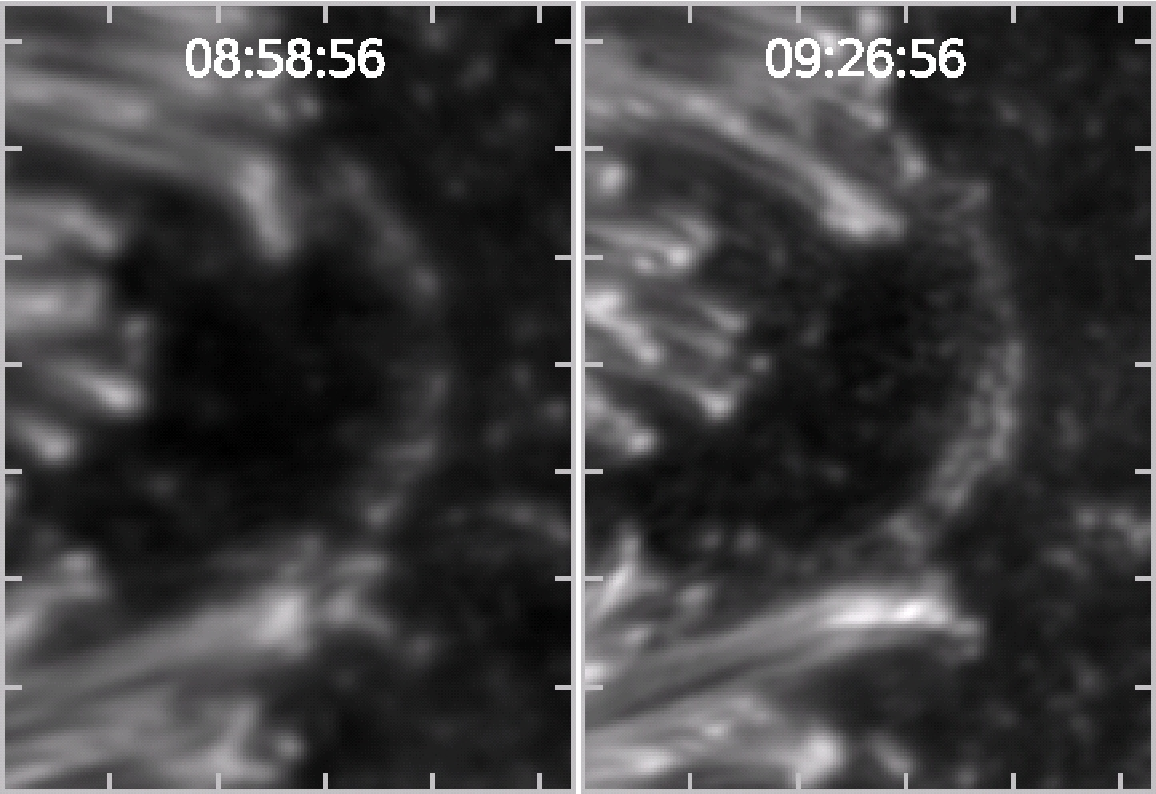}
   \caption{Most prominent chain of umbral dots in the studied umbra, located near label A in
   Fig.~\ref{FigBestImage} (FOV is 5.4~$\times$~7.4~Mm). The left panel shows the chain composed of
   several typical UDs along a curved line at 08:58:56 UT. The right panel shows the same chain
   28~min later. At this later time the lower half of the chain looks similar to a narrow light bridge.
   The typical dark lane is clearly recognizable.}
   \label{FigUdChainVersions}
   \end{figure}

   The data clearly show that bright UDs often form chains. The most prominent chains often start
   from a penumbral filament and UDs are found to move along the chain until they dissolve. The
   horizontal motion of the UDs is preferentially along the chain and is directed from the ends
   of the chain to its center, where the UDs disappear. Fig.~\ref{FigUdChainVersions} displays the
   most prominent chain of UDs in the observed umbra. The length of this chain is about 3200~km and
   it is about 350~km wide. In the left panel, from 08:58:56 UT, the chain appears as a simple
   succession of UDs. One can see the same region in the right panel 28~minutes later. The appearance
   of the chain changed and now the lower part of the chain looks similar to a narrow light bridge.
   The typical dark lane of a LB can be seen clearly, even if we degrade the image quality to the
   lower level of the left panel by convoluting the image with a point spread function of a circular
   pupil (not shown). This degradation was carried out to compensate for the higher spatial resolution
   of the later image. High spatial resolution is demonstrated by the presence of dark-cored penumbral
   filaments \citep[see][]{Langhans2007}, which are generally observed only at the highest resolution
   in the blue. The presence of the dark lane is an indication that the chain eventually evolved into
   a fully developed LB some hours after the end of the recording \citep{Katsukawa2007}.

\begin{figure}
   \centering
   \includegraphics[width=\linewidth]{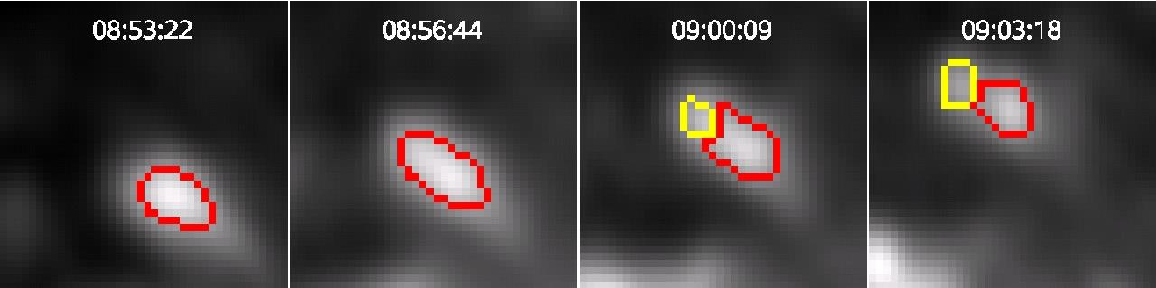}
   \caption{Splitting of an umbral dot near the position B in Fig.~\ref{FigBestImage}
   (FOV is 1.2~$\times$~1.2~Mm). The red and yellow lines are the UD boundaries as detected by the
   method explained in the main text.}
   \label{FigUdSplit}
   \end{figure}

\begin{figure}
   \centering
   \includegraphics[width=\linewidth]{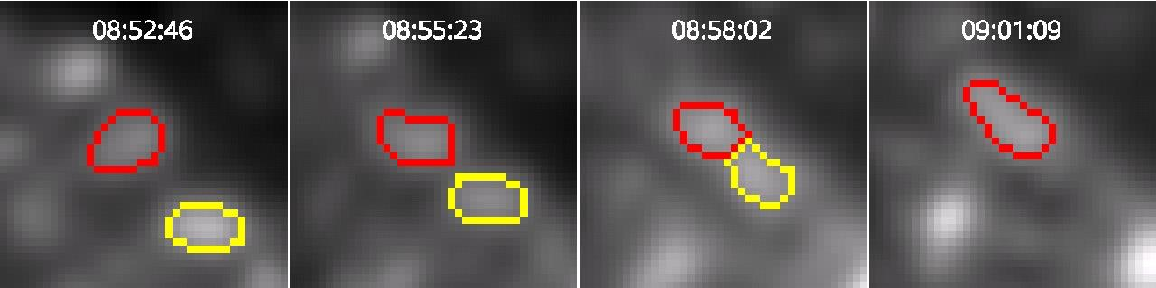}
   \caption{Merging of two umbral dots near position C in Fig.~\ref{FigBestImage}
   (FOV is 1.2~$\times$~1.2~Mm).}
   \label{FigUdMerge}
   \end{figure}

   In some exceptional cases we observe the splitting of a single UD into two parts that continue their
   life as independent UDs. Such a splitting is displayed in Fig.~\ref{FigUdSplit}. It concerns a UD
   located close to the penumbra, near point B in Fig.~\ref{FigBestImage}. The
   opposite case, the merging of two UDs into a single one, can also be observed in a few rare cases:
   See Fig.~\ref{FigUdMerge} for an example. Nearly identical phenomena in the temporal evolution of
   penumbral grains were observed by \citet{Hirzberger2002}.

\begin{figure}
   \centering
   \includegraphics[width=\linewidth]{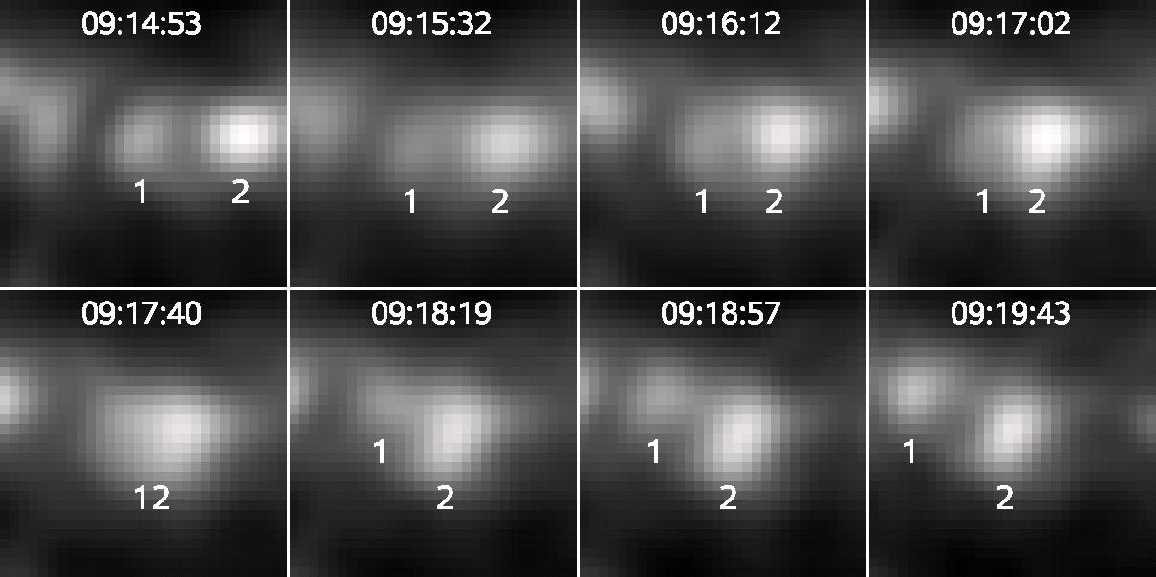}
   \caption{Elastic collision of two umbral dots. The sequence of events is seen near the position marked
   D in Fig.~\ref{FigBestImage} (FOV is 0.9~$\times$~0.9~Mm). UD~2 moves towards UD~1
   and merges with it, before separating from it again.}
   \label{FigUdNudge}
   \end{figure}

   An interesting sequence of events is illustrated in Fig.~\ref{FigUdNudge}. First the merging of two
   UDs can be observed followed immediately afterwards by the splitting of the resulting, unified UD
   into two new UDs. The sequence looks similar to an elastic impact in a Newton pendulum consisting of
   two balls: The nearly motionless UD~1 is hit by the moving UD~2. The impact of the two UDs brings
   the second UD to a standstill while the first UD starts to move roughly in the direction of the first UD.
   Note that this is simply an empirical description and we do not propose that this is what physically
   happens (or that the unified UD breaks up into the same parcels of gas which united to form it).

   If two UDs come close to each other then the visual impression of a single UD exhibiting a
   dark lane can occur, see last panel of Fig.~\ref{FigUdNudge} for an example. None of the UDs in our data
   set seems to stay in such a state for a significant fraction of the UD lifetime. Since the dark lanes as
   seen in the simulations of \citet{Schuessler2006} are visible for most of the UD lifetime, we conclude
   that we find no clear evidence for such dark lanes. This difference to the results of \citet{Bharti2007}
   and \citet{Rimmele2008} may stem from the different wavelengths of the analyzed data. The wavelength can
   have a remarkable effect on the detected fine structure \citep[e.g.][]{Zakharov2008}.

   The properties of a UD change along its trajectory. In order to assign a property to the entire
   trajectory we either average over all points along the trajectory (which is expressed by
   introducing an upper index "Mean") or we determine the maximum value reached by that parameter
   over all trajectory points (which is expressed by the upper index "Max"), e.g., $D^{Mean}$ means
   the UD diameter averaged along the trajectory, while $I_{Peak}^{Max}$ is the maximum value of all
   peak intensities along a UD trajectory (where the peak intensity is the largest intensity within
   the UD boundary at a given point in time).

\begin{figure}
   \centering
   \includegraphics[width=0.5\linewidth-1mm]{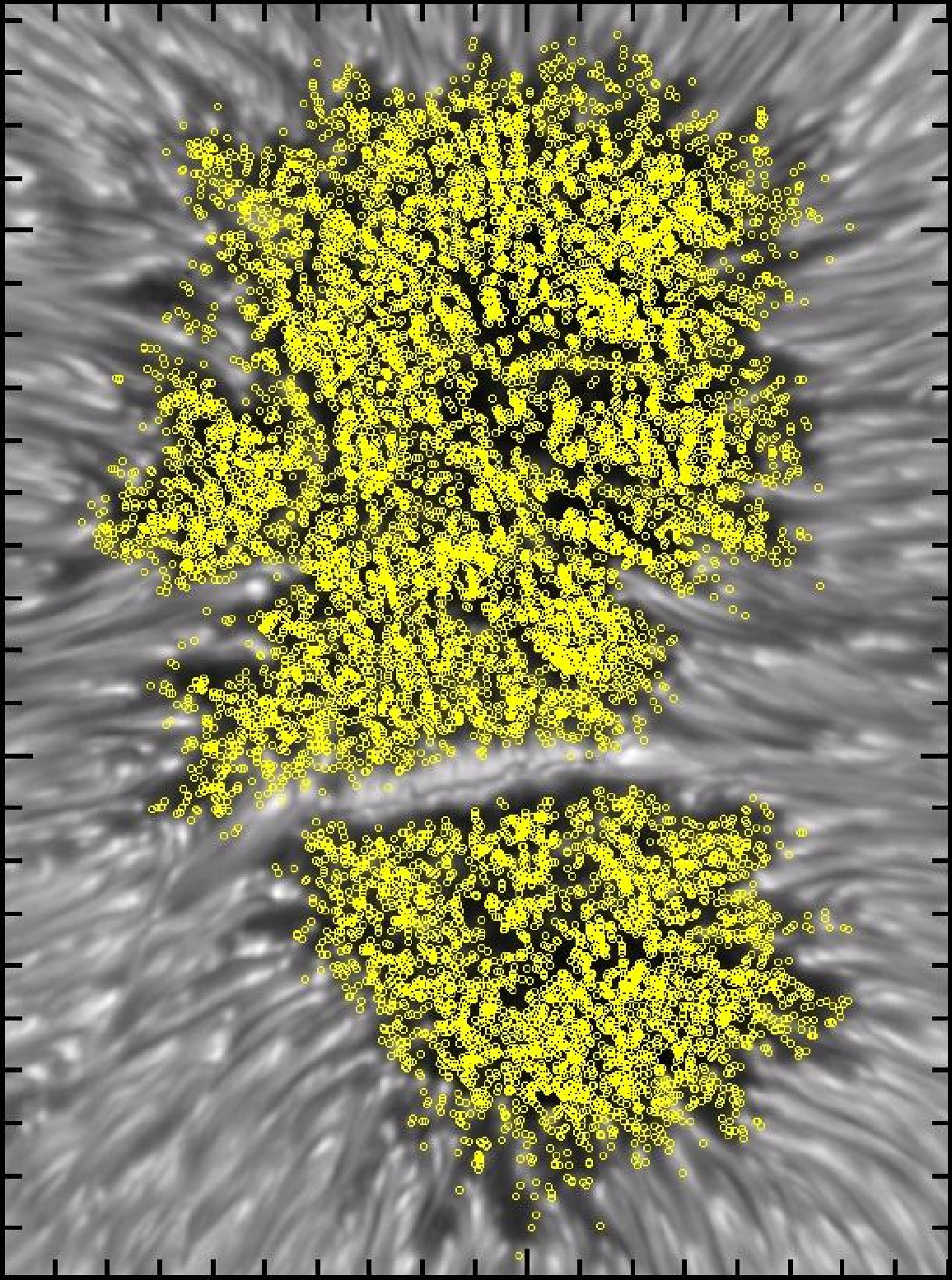}
   \includegraphics[width=0.5\linewidth-1mm]{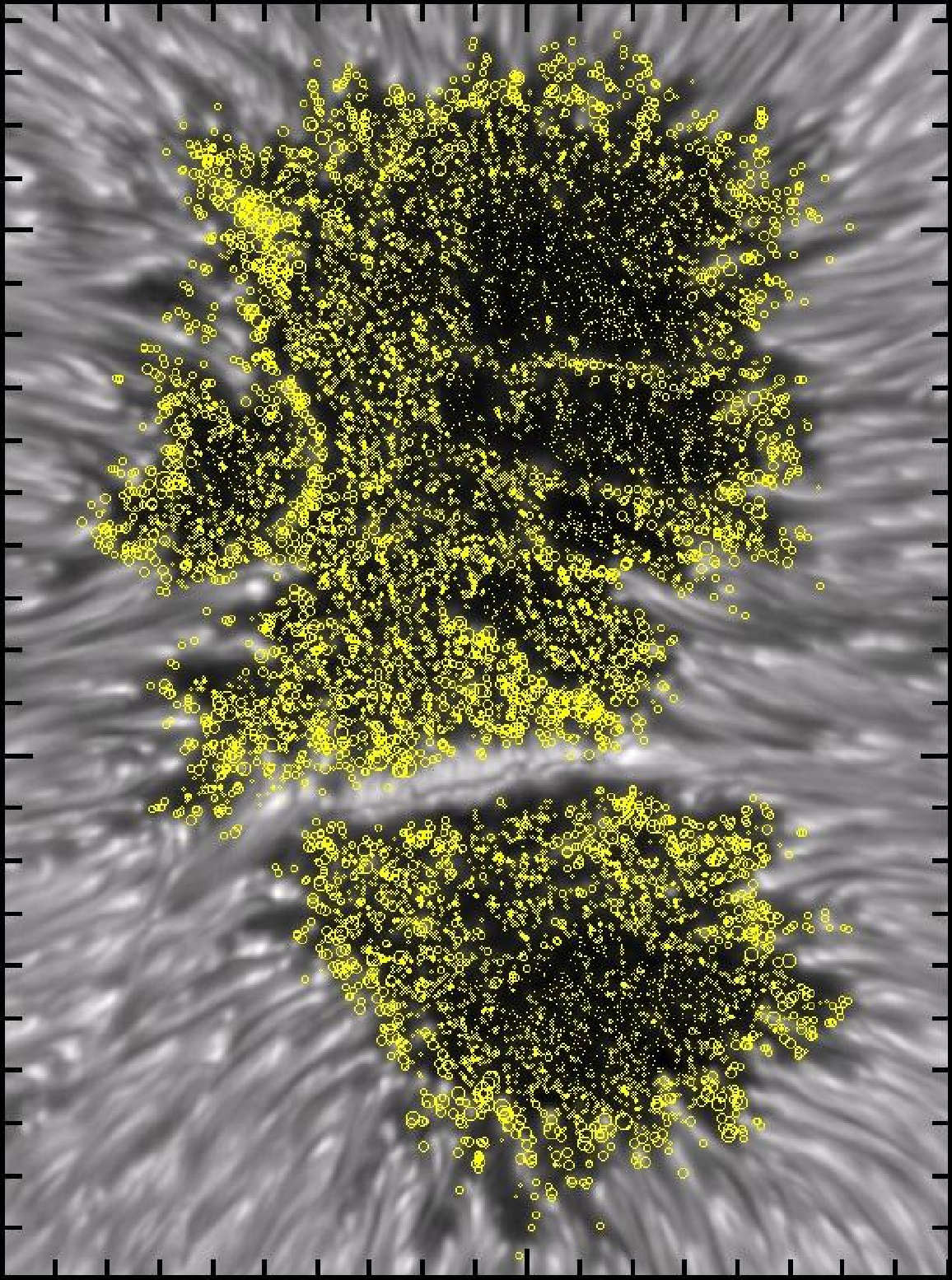}
   \caption{Spatial distribution of UD occurrence (left panel) and UD maximum brightness (right panel).
   Each UD is plotted only once along its entire trajectory, at the position of its maximum brightness.}
   \label{FigUdOverview}
   \end{figure}

   Fig.~\ref{FigUdOverview} shows the best image with all identified UDs marked by circles. The
   circles are centered on the positions of $I_{Peak}^{Max}$, i.e. the position of maximum intensity.
   In the left panel the circles have a constant radius, while their radii are proportional to $I_{Peak}^{Max}$
   in the right panel. As a result we get an impression of the spatial distribution of UD occurrence
   as well as of the spatial distribution of UD brightness. Obviously, there is hardly any part of
   the umbra which does not support umbral dots. Only very localized small voids are visible.
   The brightness distribution of UDs, however, is rather inhomogeneous, with clear concentrations
   of bright UDs and regions harboring mainly dark UDs (mainly in the upper right part of the
   upper umbra and in the lower part of the lower one). Since the UDs of a chain (like the chain close
   to label A in Fig.~\ref{FigBestImage} and shown in more detail in Fig.~\ref{FigUdChainVersions})
   appear to move along the chain and hardly in the direction perpendicular to it, there is often a narrow
   void directly beside the chain.

   \subsection{Quantitative properties}

\begin{figure}
   \centering
   \includegraphics[width=\linewidth-1mm]{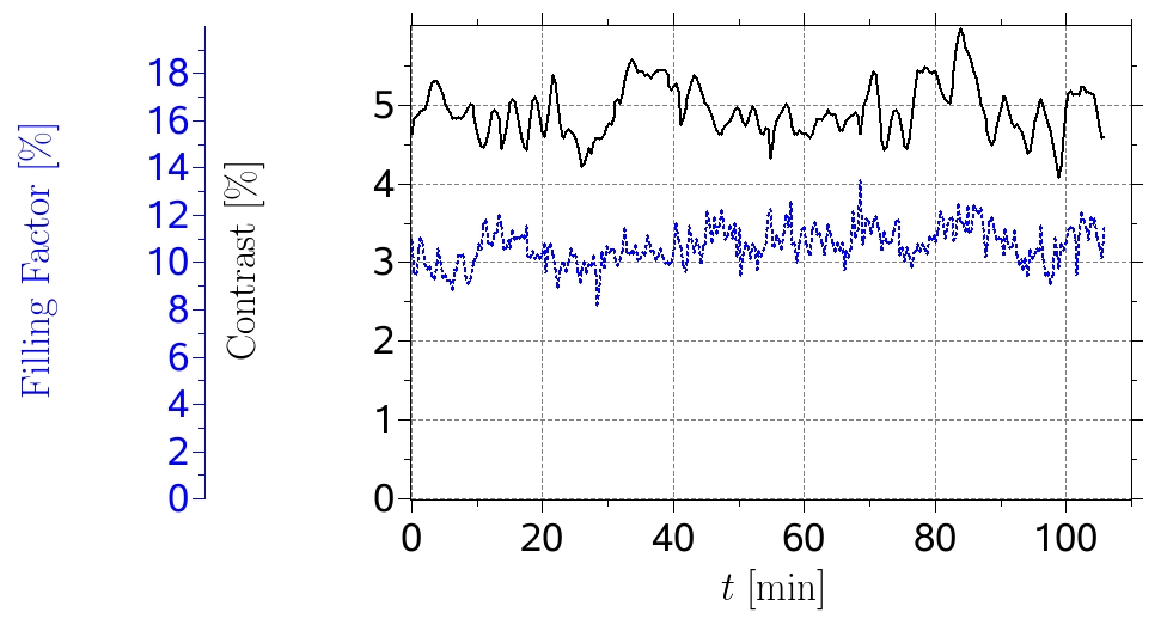}
   \caption{Dependence of UD filling factor (dotted blue line) and granulation contrast
   (solid black line) on the interval of time, $t$, since the first image of the series was recorded.}
   \label{FigFillingFactor}
   \end{figure}

   The filling factor, i.e. the sum of all UD areas relative to the total umbral area, is correlated
   to the image quality. Essentially, the filling factor is constant over the entire period of
   observations (see Fig.~\ref{FigFillingFactor}), which is important for the later determination of the
   time dependence of their properties. On average we determine a value of about 11\,\%. Fig.~\ref{FigFillingFactor}
   shows also the image contrast from an undisturbed granulation area outside the sunspot which demonstrates the
   high homogeneity of the image quality in our time series.

\begin{figure}
   \centering
   \includegraphics[width=\linewidth-1mm]{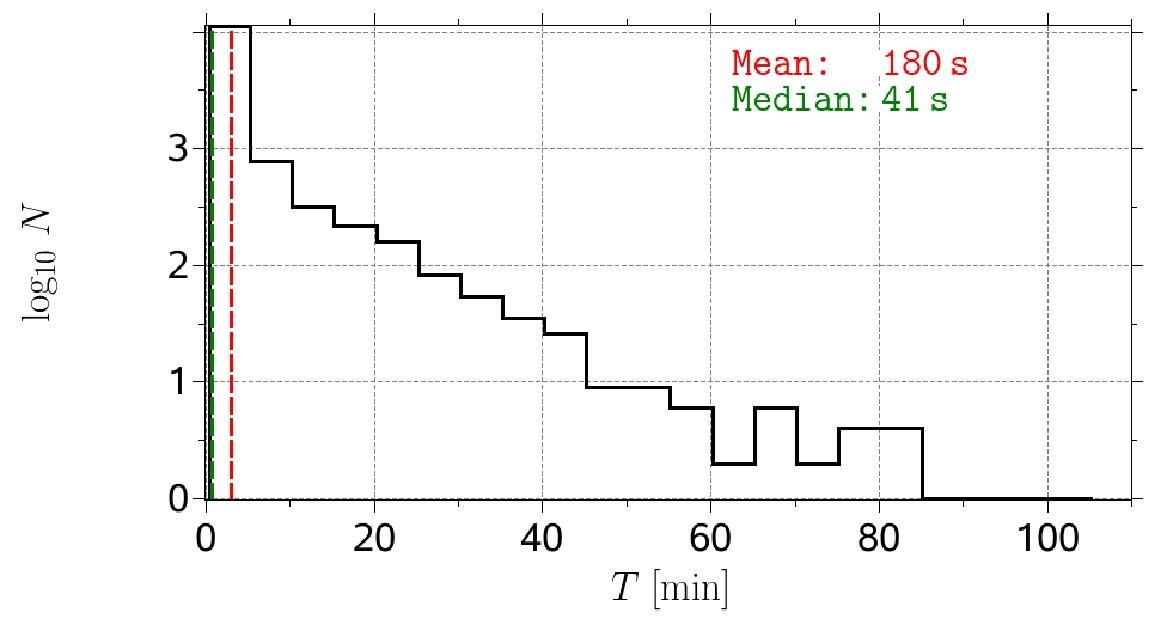}
   \caption{Histogram of lifetimes $T$ of all 12836 UD trajectories (solid line) and their mean and
   median value (dashed lines and text labels). The bin size is 300~s.}
   \label{FigHistLifetime}
   \end{figure}

   The histogram of the UD lifetimes is displayed in Fig.~\ref{FigHistLifetime} with a logarithmic $y$ axis.
   One can see that most of the UDs live for a short time. These short-lived UDs move over
   short distances which leads to physically nonsensical velocities due to the discretization of lifetime
   and distance. Consequently, when discussing trajectories
   and velocities of UDs we only consider the 2899 trajectories of UDs with lifetimes greater than 150~s.
   The histogram is nearly linear for lifetimes between 5 and 60 minutes, which, due to the logarithmic
   vertical scale, suggests an exponential distribution of lifetimes. The excess of UDs with short
   lifetimes may partly be due to seeing. Also given in Fig.~\ref{FigHistLifetime} is the mean lifetime
   (180~s) and the median lifetime (41~s). If we consider only the 2899 trajectories of UDs with lifetimes
   greater than 150~s then we find a mean lifetime of 630~s and a median of 390~s. Note that 281 UDs are
   already present in the first image and 344 UDs are still present in the last image, whereas
   only one UD survives the whole sequence. Ignoring those 625 UDs reduces the mean lifetime
   from 180~s to 152~s. The median lifetime as well as the shape of the histogram do not change, because
   the number of incomplete UD trajectories is small compared to the total number of UDs. Thus we decided
   to neglect this effect and consider all 12836 UDs in the following text.

\begin{figure*}
   \centering
   \includegraphics[width=0.5\linewidth-4mm]{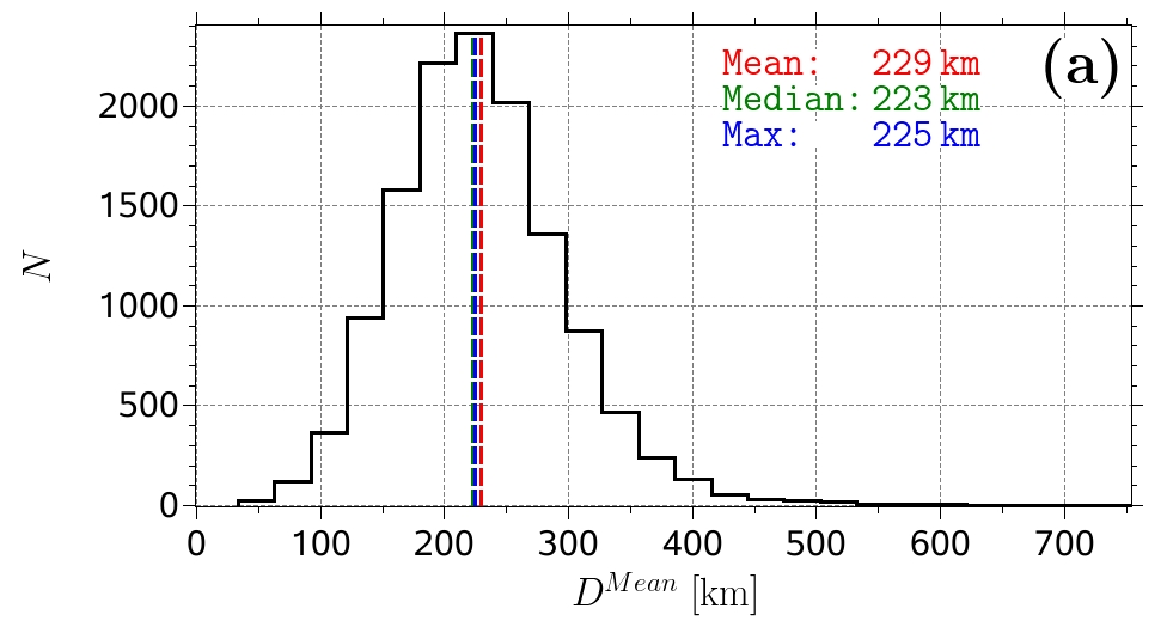}
   \hspace{5mm}
   \includegraphics[width=0.5\linewidth-4mm]{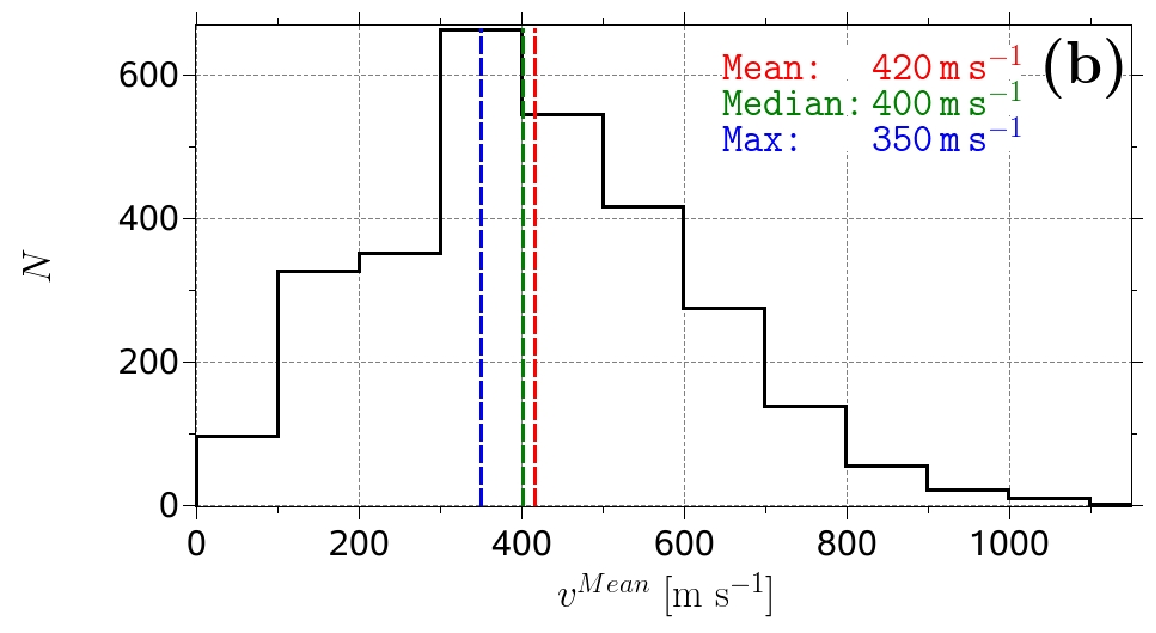}\\[5mm]
   \includegraphics[width=0.5\linewidth-4mm]{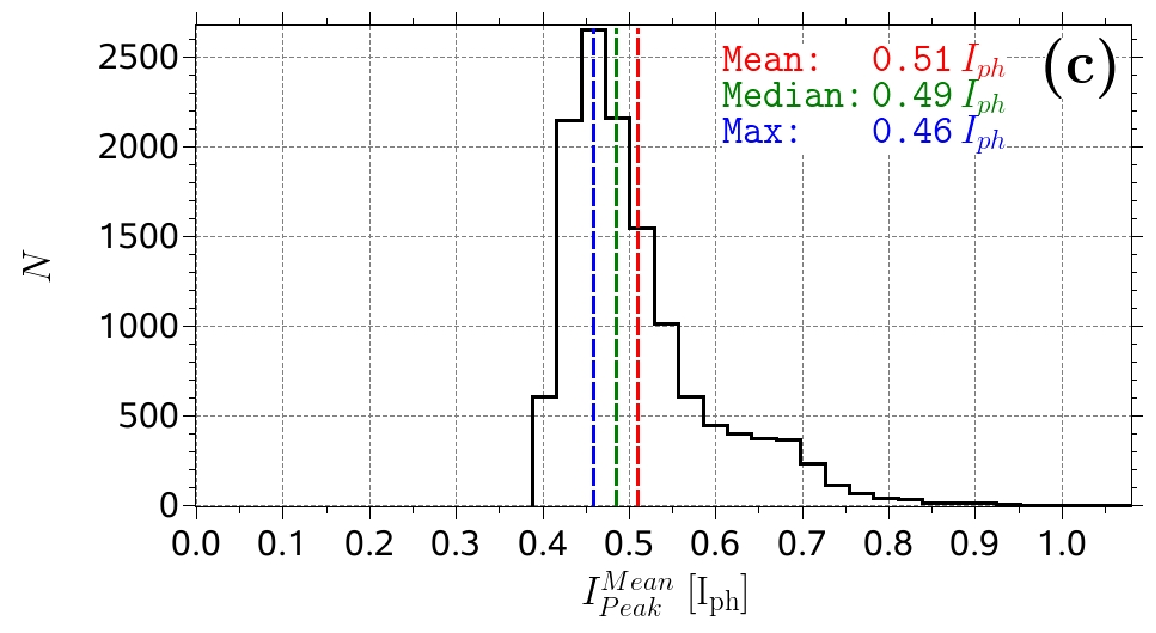}
   \hspace{5mm}
   \includegraphics[width=0.5\linewidth-4mm]{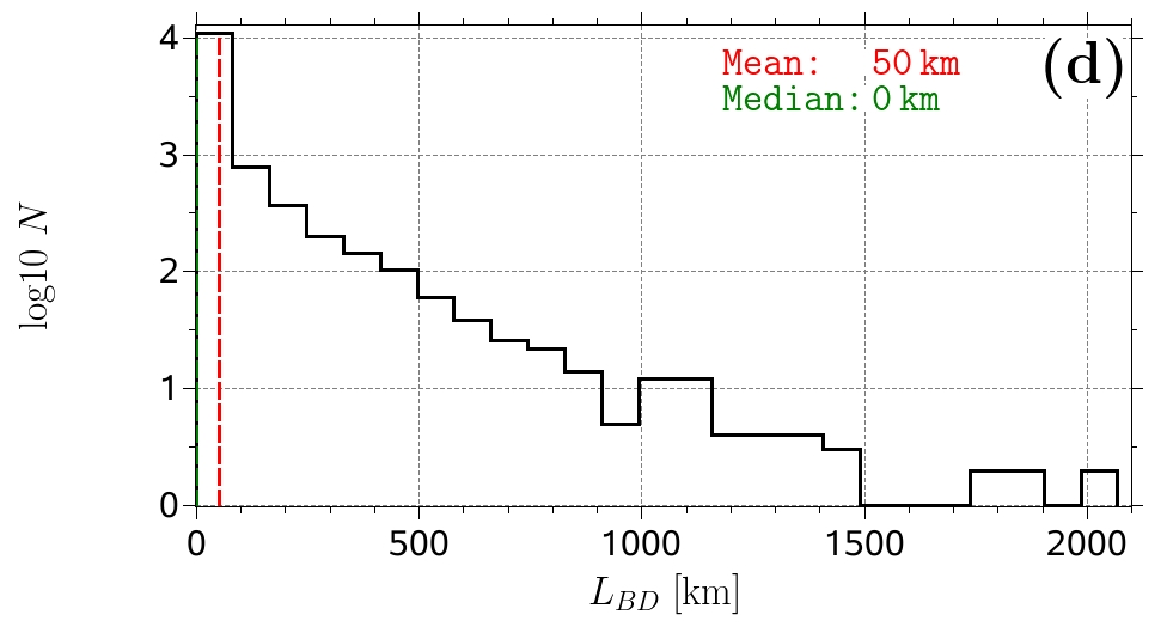}
   \caption{Histogram of mean diameters (a), mean horizontal velocities (b), mean peak intensities (c), and
   distances between birth and death position (d). (a), (c), and (d) are plotted for all 12836 UD trajectories
   and (b) for the 2899 trajectories of UDs that lived longer than 150~s. The location of the
   maximum, the mean, and the median of the distribution is indicated in each frame. The bin sizes
   are 30~km for (a), 100~m\,s$^{-1}$ for (b), 0.03~$I_{ph}$ for (c), and 90~km for (d).}
   \label{FigHistOverview}
   \end{figure*}

\begin{figure}
   \centering
   \includegraphics[width=\linewidth-1mm]{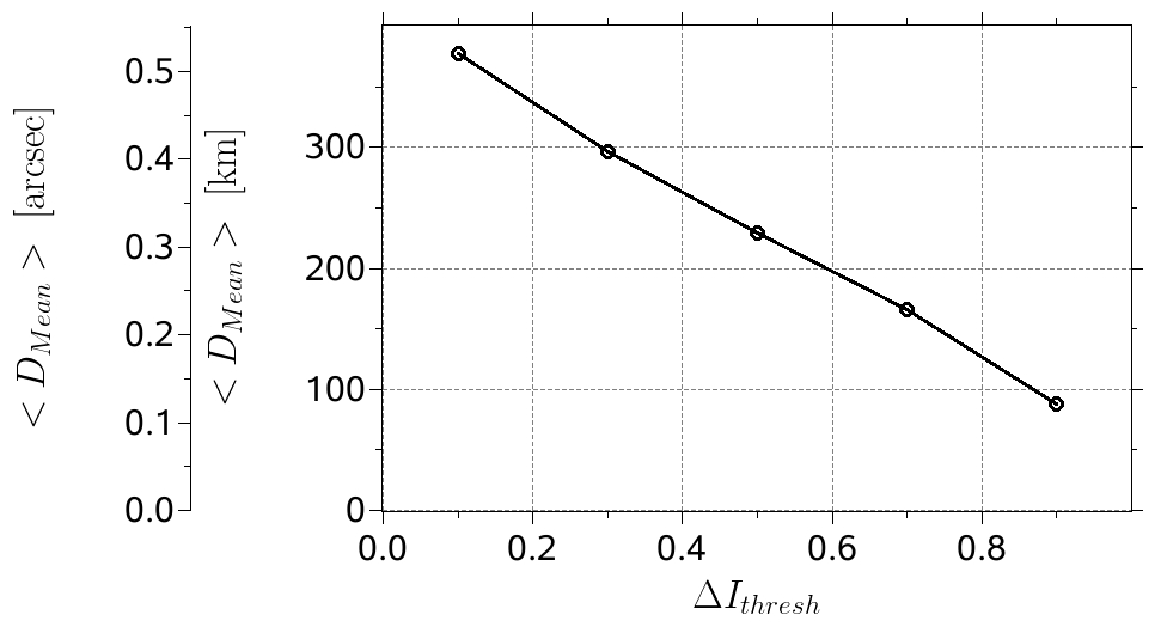}
   \caption{Mean UD diameter averaged over all trajectories as a function of the intensity threshold
   $\Delta I_{thresh}$ (defined in main text) that is used to determine the UD boundaries
   (see also Fig.~\ref{FigCuttingPixels} and its explanation).}
   \label{FigDiameterVsBoundThresh}
   \end{figure}

   The histogram of the mean UD diameters is plotted in Fig.~\ref{FigHistOverview}~a.
   The mean UD diameters vary between 50 and 750~km. The UDs have a mean diameter of around
   229~km and 95\,\% of the UDs are spatially resolved at our diffraction limit of 130~km.
   \citet{Sobotka2005} found 175~km for the mean UD diameter. They determined
   the UD boundaries by finding all pixels with downward concavity, whereas we used all pixels whose
   intensity is greater than 50\,\% of the $I_{Peak}-I_{bg}$ range,
   $\Delta I_{thresh}=(I_{thresh}-I_{bg})/(I_{Peak}-I_{bg})=0.5$. As one can see in
   Fig.~\ref{FigDiameterVsBoundThresh} we would also find a mean UD diameter of 175~km if a
   threshold of about $\Delta I_{thresh}=0.67$ would be used. Nevertheless, we prefer to continue
   our analysis with a value of 0.5 in analogy to the FWHM definition. Clearly, the employed threshold
   influences the filling factor as well, roughly quadratically. Remarkably, the histogram is nearly
   symmetric, which supports the conclusion that most UDs have been resolved.

   The histogram of the mean horizontal velocities, i.e. the quotient of trajectory length and lifetime,
   is plotted in Fig.~\ref{FigHistOverview}~b and exhibits a broad distribution from 0 to more than
   1~km\,s$^{-1}$ with a significant maximum at 350~m\,s$^{-1}$. The velocity distribution is slightly
   asymmetric with a small tail to higher velocities. Fig.~\ref{FigHistOverview}~c shows the histogram of
   the mean peak intensities, i.e. the mean of all peak intensities of the points along the trajectory.
   All intensities are normalized to the mean intensity of the quiet photosphere ($I_{ph}$). Just 3
   of the 12836 UDs reach a brightness greater than that of the quiet Sun, whereas most of the UDs
   are about half as bright as the quiet photosphere. The distribution is asymmetric, with a tail to
   higher intensities. Note that these brightnesses are strongly wavelength dependent and cannot
   be easily compared with values published in the literature \citep{Solanki2003}. Lastly, the
   histogram of the distances between the UD's birth and death position ($L_{BD}$) is given in
   Fig.~\ref{FigHistOverview}~d. With increasing distance the number of UDs decreases exponentially,
   so that only a few UDs travel over long distances in their life, most UDs do not leave the vicinity
   of their birth position. The maximum observed birth-death distance is 2~Mm, about 20\,\% of the upper umbra's
   diameter of roughly 10~Mm.

\begin{figure}
   \centering
   \includegraphics[width=\linewidth-1mm]{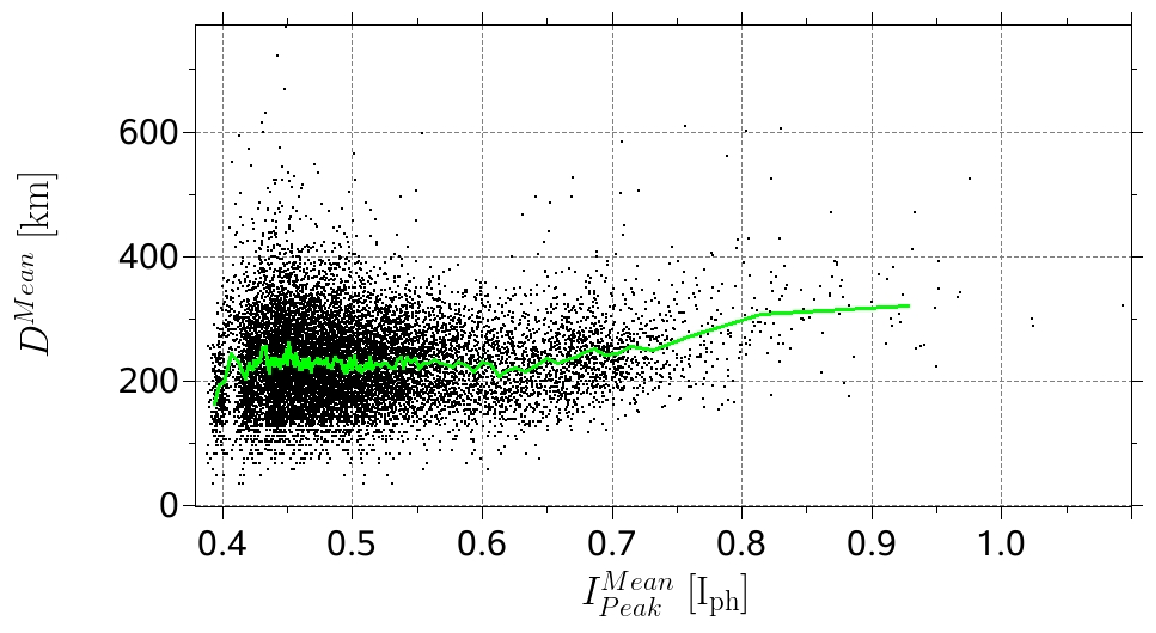}
   \caption{Scatterplot of mean UD diameter versus mean peak intensity. The solid green line connects binned values.}
   \label{FigScatterMeanDiameterVsMeanPeakIntensity}
   \end{figure}

   A UD size versus UD peak intensity scatterplot (Fig.~\ref{FigScatterMeanDiameterVsMeanPeakIntensity})
   reveals that there is a weak correlation between UD size and brightness. The brightest UDs are
   not the biggest ones and large UDs are not the brightest ones. The solid green line connects binned values
   (obtained by averaging 100 data points with similar intensities) and shows that bright UDs are on
   average a bit larger than small ones. The relation found by \citet{Tritschler2002} is qualitatively
   confirmed, although they only considered UD intensities in individual snapshots while we tracked the
   temporal development of the UDs over their lifetimes.

\begin{figure}
   \centering
   \includegraphics[width=\linewidth-1mm]{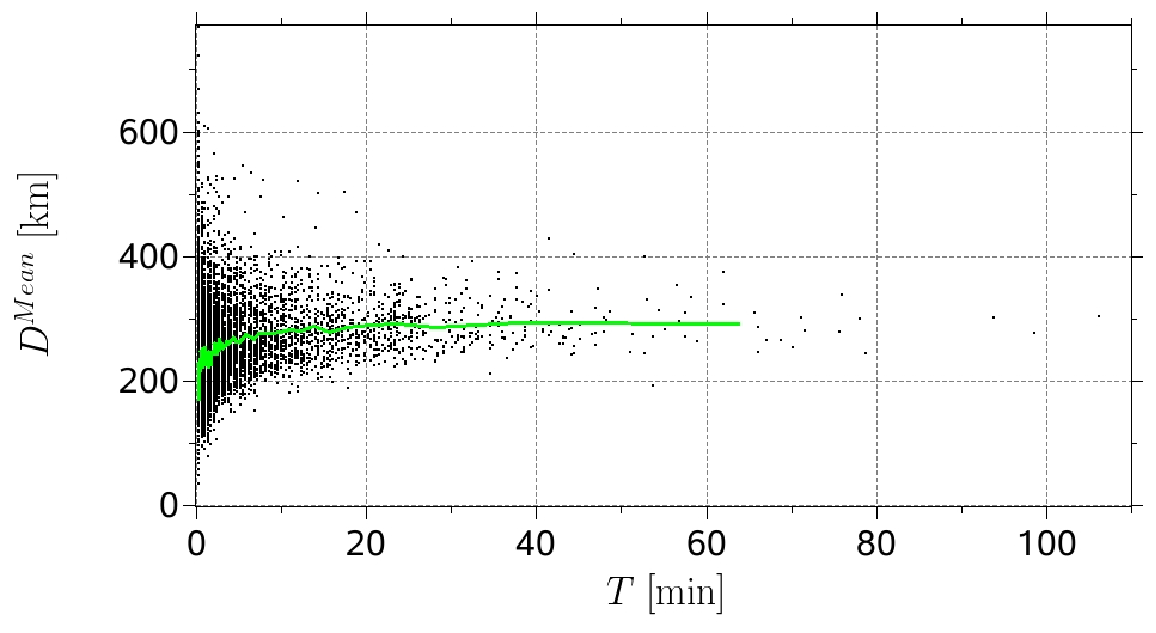}
   \caption{Scatterplot of mean UD size versus lifetime. The solid green line connects binned values.}
   \label{FigScatterMeanDiameterVsLifetime}
   \end{figure}

   The relation between a UD's mean size (i.e. the size averaged over the lifetime) and its lifetime is
   plotted in Fig.~\ref{FigScatterMeanDiameterVsLifetime} (the green curve is obtained after binning over
   100 data points). The binned values show an increase in $D^{Mean}$ with increasing lifetimes for
   short-lived UDs. For the longer lived ones size and lifetime do not correlate. The UD sizes scatter
   more for short lifetimes. All long-lived UDs are of intermediate size of around 290~km. The large,
   short-lived UDs are all present in the first image of the time series, so that their lifetime would
   actually be larger if we had started our observation earlier.

   In the literature we often find a separation into two UD classes, e.g. \citet{Grossmann1986}
   find a difference between peripheral and central UDs, i.e. between UDs that are born
   close to the umbra-penumbra boundary and UDs that are born deep in the umbra, whereas
   \citet{Hartkorn2003} and \citet{Sobotka1997b} distinguish bright and dark UDs. We use
   different properties to find reasonable distinctions between types of UDs. E.g. from now on
   a given UD is called a peripheral UD (PUD) if the UD's birth position is closer than 400~km
   to the umbral boundary, otherwise it is termed a central UD (CUD). (The selected
   threshold comes from the histogram of the distances between the UD's birth position and the
   umbral boundary (not shown) which shows a maximum at around 400~km.) Alternatively, if the
   distance traveled between birth and death position is larger than 750~km then we call it
   a mobile UD, otherwise a stationary UD. (We plotted the trajectories of all UDs
   whose $L_{BD}$ was greater than a threshold, which was determined by starting at a
   small value and increasing it step by step. We stopped at 750~km which is the
   smallest $L_{BD}$ at which no trajectories occurred anymore in the central part of the umbra.)
   The aim here is not to separate UDs into distinct
   classes by a single property, e.g. a histogram of $L_{BD}$ (Fig.~\ref{FigHistOverview}~d)
   does not show two peaks, even if restricted to long-lived UDs. However, as we shall see below
   the most mobile UDs are formed near the penumbra, while the least mobile ones are mainly
   formed deep in the umbra. Such a distinction may help to guide theory towards a better
   understanding of the origin and evolution of UDs with different properties and at different
   locations.

   \begin{figure*}
   \centering
   \includegraphics[width=0.5\linewidth-1mm]{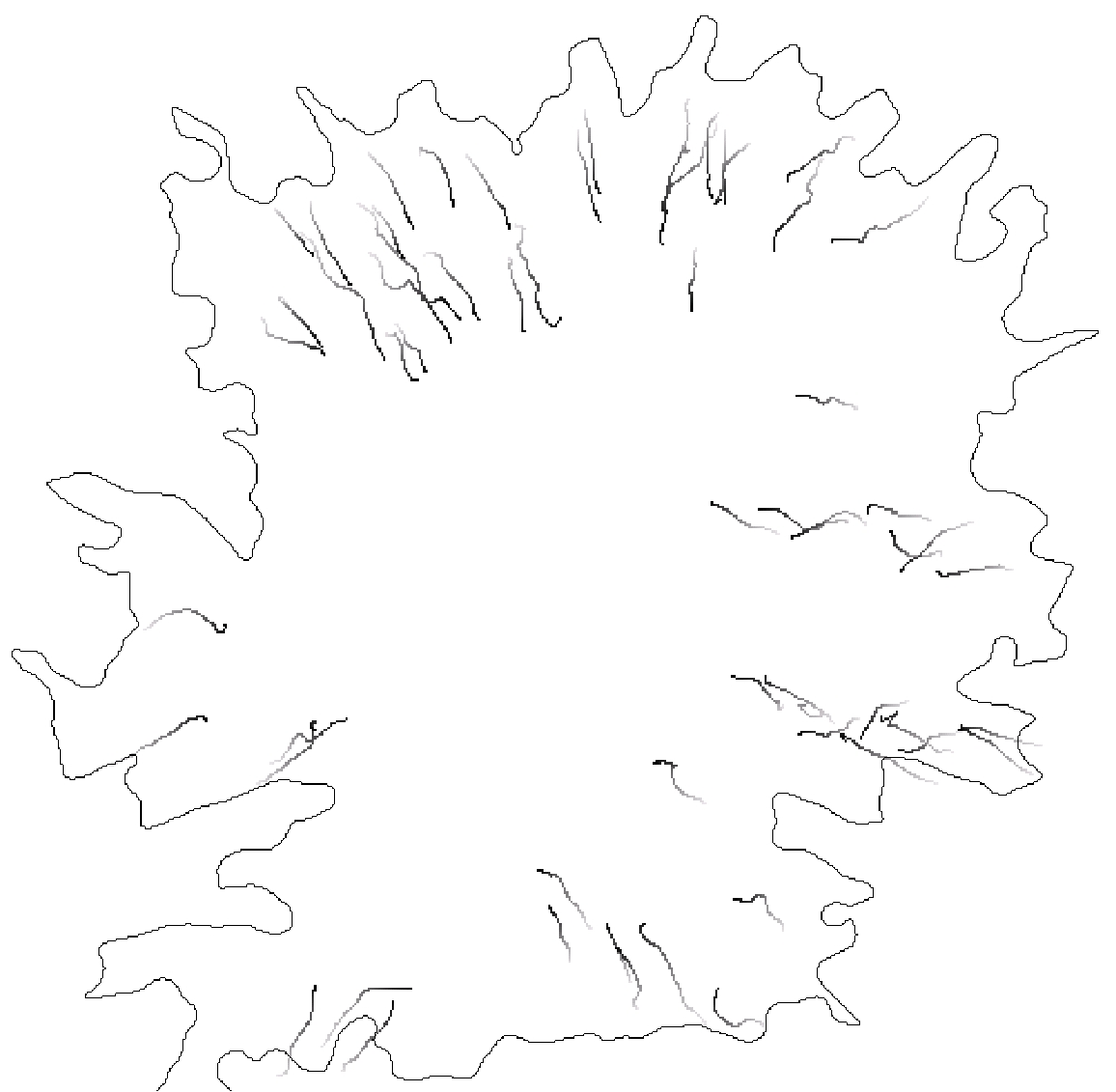}
   \includegraphics[width=0.5\linewidth-1mm]{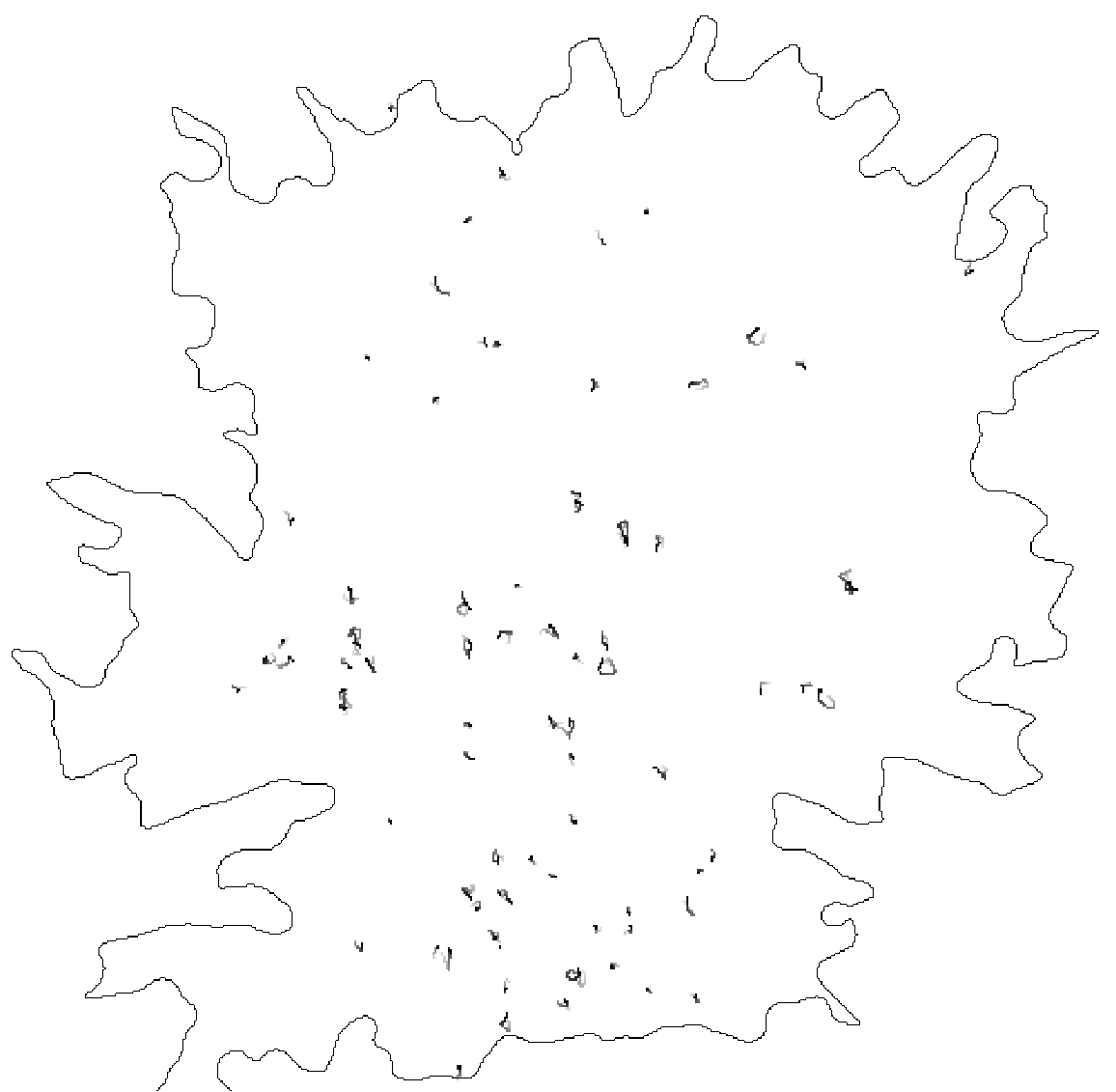}
   \caption{UD trajectories in the upper umbra. The bright ends of the trajectories denote the positions
   of the UD's birth and the dark ends show the position at death. The black contour line corresponds to the
   umbral boundary as detected in the best quality image. The left panel shows all UDs whose distance between
   the birth and the death position was greater than 750~km, the right panel shows all UDs with a birth-death
   distance smaller than 150~km whose lifetime was greater than 800~s.}
   \label{FigUdTrajectoriesBirthDeathDist}
   \end{figure*}

\begin{figure*}
   \centering
   \includegraphics[width=0.5\linewidth-1mm]{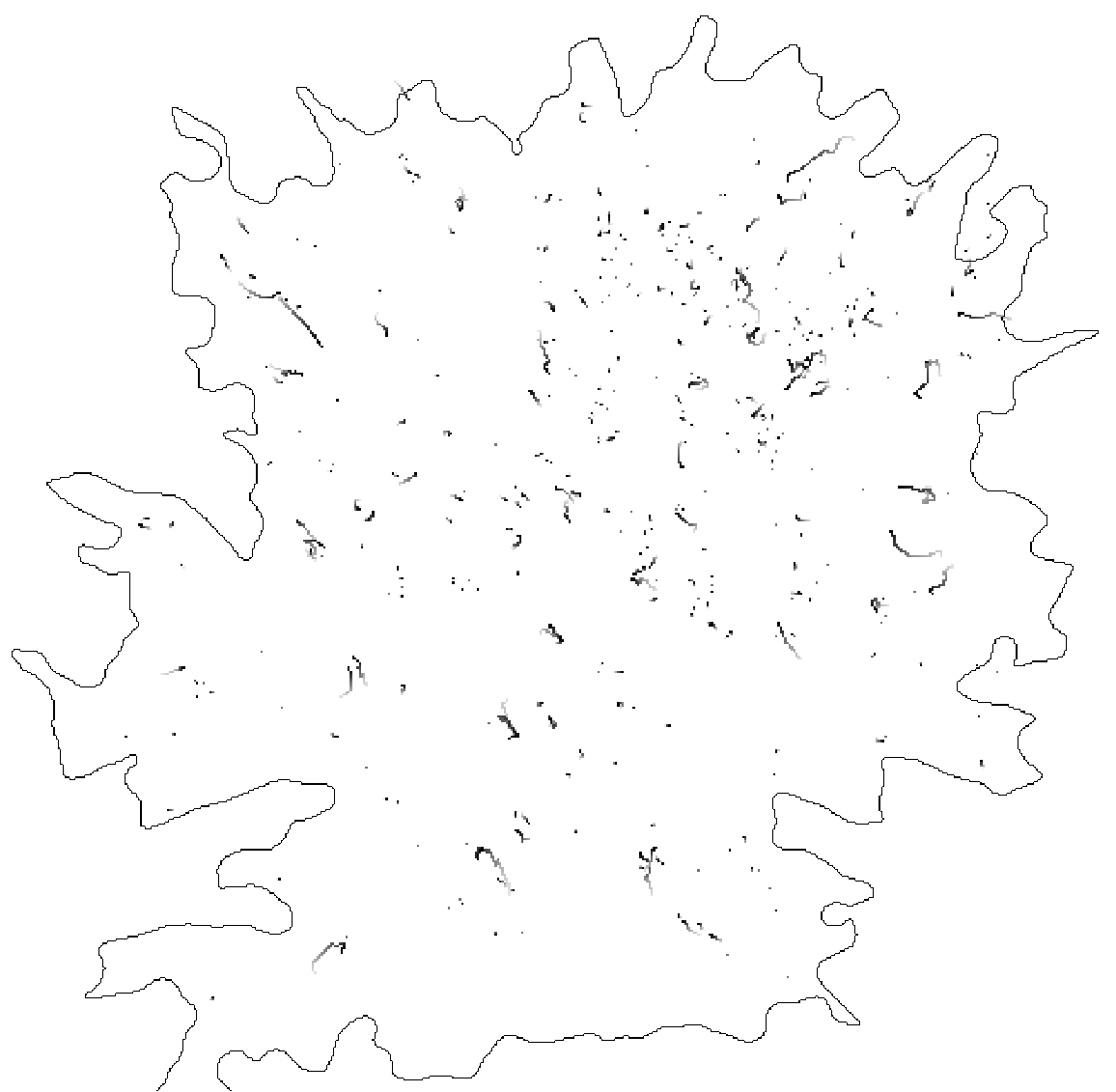}
   \includegraphics[width=0.5\linewidth-1mm]{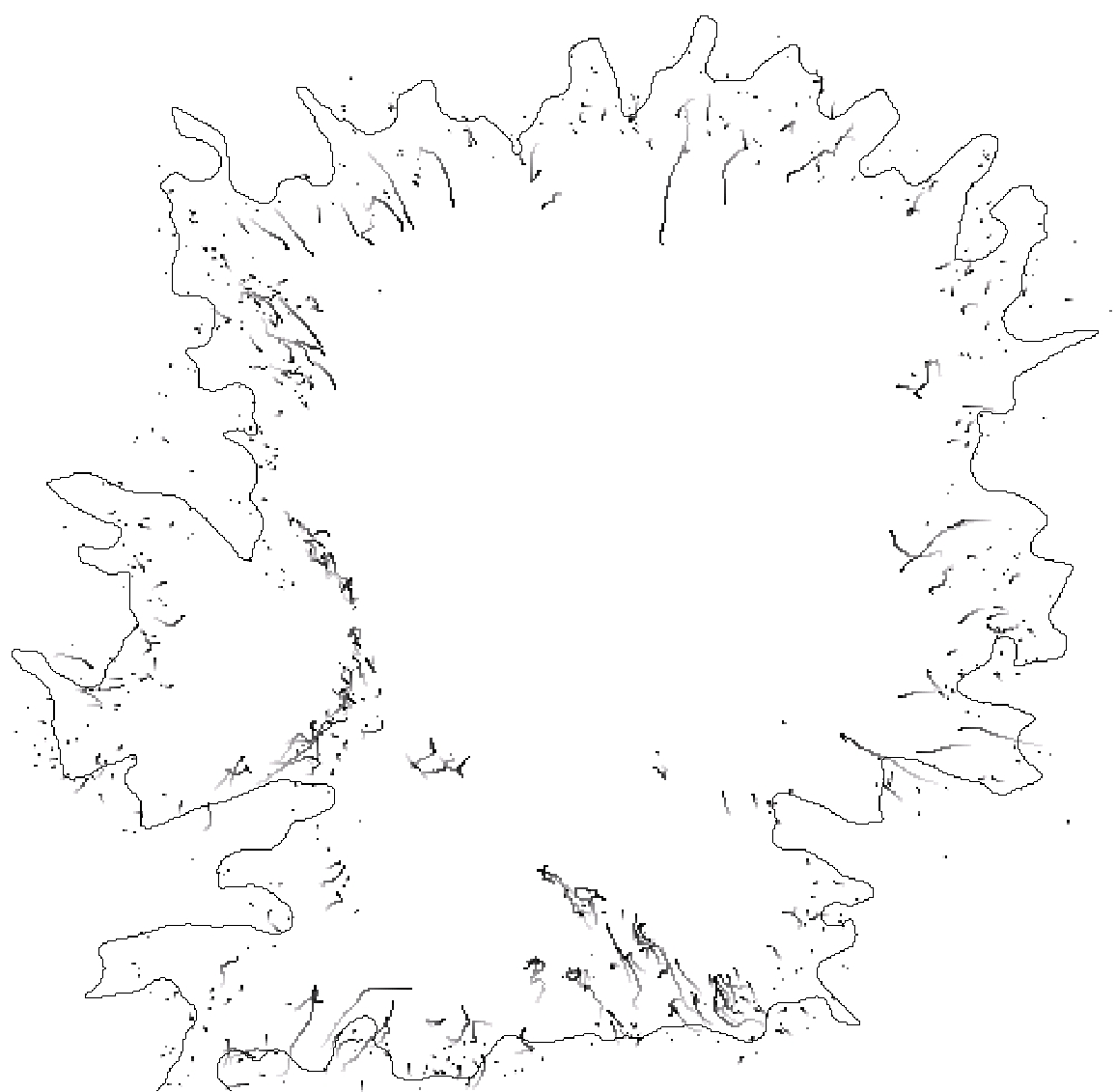}
   \caption{UD trajectories in the upper umbra. The left panel shows all UDs with mean diameter
   greater than 350~km and the right panel shows all UDs with a mean peak intensity greater than 0.65~$I_{ph}$.}
   \label{FigUdTrajectoriesDiameterAndPeakIntensity}
   \end{figure*}

\begin{figure}
   \centering
   \includegraphics[width=\linewidth-1mm]{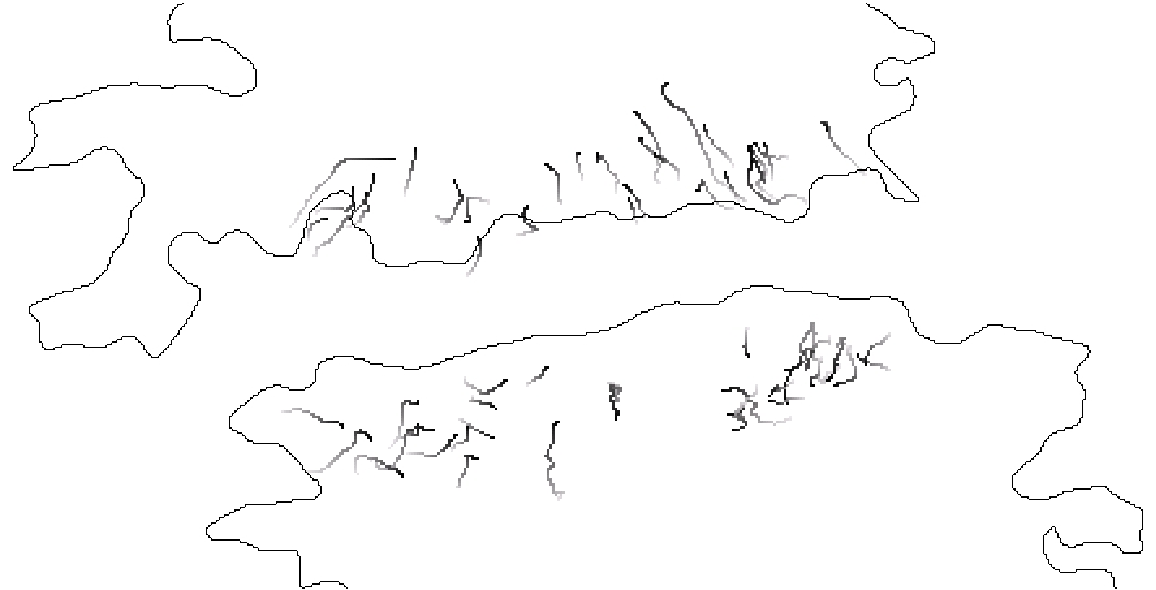}
   \caption{UD trajectories near the light bridge with birth-death distance greater than 300~km.}
   \label{FigLightbridge}
   \end{figure}

   The strong concentration of bright UDs near the umbral boundary (or along proto-light bridges)
   is already clear from Fig.~\ref{FigUdOverview} (right panel). Fig.~\ref{FigUdTrajectoriesBirthDeathDist}
   shows a separation by birth-death distance. The left panel displays the longest UD trajectories, i.e.
   only the mobile UDs are shown. These UDs are also relatively long-lived. (Smallest lifetime of this UD
   class, that contains 85 UDs, is 15~min.) The right panel shows trajectories of long-lived UDs with a
   small birth-death distance. A clear separation by the birth position is readily identifiable. Almost
   all trajectories with a large birth-death distance start close to the umbra-penumbra boundary
   and these UDs move nearly radially into the umbra. A visual inspection of the movie of the reconstructed
   time series of images shows that many of these UDs are former penumbral grains that broke away from the
   penumbra. In contrast, many of the UDs with a small birth-death distance are born in
   the umbral interior and move along a closed loop or jitter around their birth position. We cannot say
   if this jitter has a physical background or if it is caused by residual seeing-induced noise. As mentioned
   in section~3 we determined the umbral boundary individually for each image but we show only that
   corresponding to the best quality image as black contour line in Fig.~\ref{FigUdTrajectoriesBirthDeathDist},
   \ref{FigUdTrajectoriesDiameterAndPeakIntensity}, and \ref{FigLightbridge}. Consequently some trajectories
   (or parts of them) are outside the black contour line, although they are always inside the umbra at the time
   of their occurrence (see, e.g., the bottom-left corner of the left plot of
   Fig.~\ref{FigUdTrajectoriesBirthDeathDist}).

\begin{figure}
   \centering
   \includegraphics[width=\linewidth-1mm]{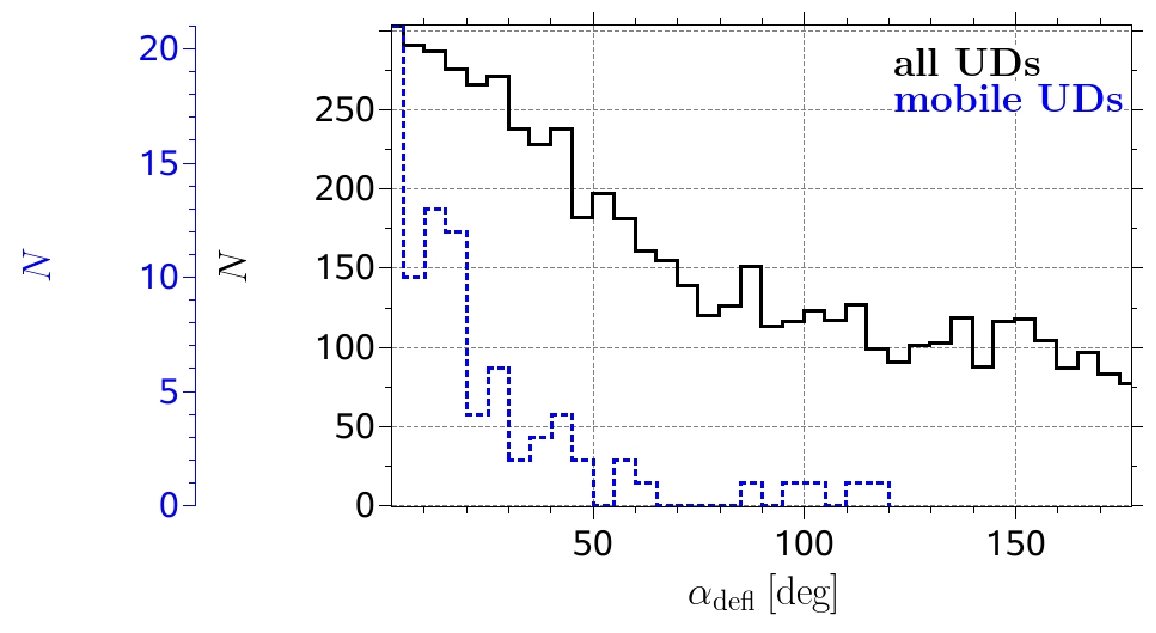}
   \caption{Histogram of radial deflection angles $\alpha_{\rm{defl}}$ of all UD trajectories (solid black line)
   and of the mobile UD trajectories (dotted blue line).}
   \label{FigHistRadialDeflection}
   \end{figure}

   Let us now consider more quantitatively the fact that the mobile UDs prefer to move radially towards the
   umbral center. Fig.~\ref{FigHistRadialDeflection} displays a histogram of the UD's deflection angles
   $\alpha_{\rm{defl}}$ that is defined as angle between the line connecting the umbral center and the UD's
   birth position and the line connecting the UD's birth and death position. Radially directed inward flow
   will lead to $\alpha_{\rm{defl}}~=~\mathrm{0^\circ}$ and an outward flow to
   $\alpha_{\rm{defl}}~=~\mathrm{180^\circ}$. Obviously, this definition makes sense only for the 5689 UD
   trajectories that have different birth and death positions. The solid black line in
   Fig.~\ref{FigHistRadialDeflection} shows the histogram for all these UDs and exhibits a clear tendency
   for a radially directed inward motion (42\,\% of the UDs are found to have a deflection angle lower than
   $\mathrm{45^\circ}$). This tendency is much more significant if we only consider mobile UDs, see the
   dotted blue line (88\,\% of the mobile UDs are found to have a deflection angle lower than
   $\mathrm{45^\circ}$). However, histograms calculated for central UDs and for peripheral UDs (not shown)
   lead, in principle, to the same shape as for all UDs, i.e. UDs born close to the penumbra do not show a
   significantly higher tendency of radially inward directed motion than the central UDs.

   The trajectories of large UDs are drawn in Fig.~\ref{FigUdTrajectoriesDiameterAndPeakIntensity} (left panel).
   They are born throughout the umbra, with a tendency to cluster in the darker part of the umbra. The vicinity
   of the light bridge is avoided. This UD class contains long trajectories as well as short ones. The
   trajectories of the brightest UDs can be seen in the right panel of
   Fig.~\ref{FigUdTrajectoriesDiameterAndPeakIntensity}. These UDs all emerge close to the penumbra, the light
   bridge, or the prominent UD chain (label A in Fig.~\ref{FigBestImage}).

   Many UDs with a preferred direction of motion arise near the light bridge. Most of the UDs that emerge
   on the disk center side of the light bridge (i.e. into the large umbra) move away from the light bridge
   while many of the limbside UDs (i.e. those formed in the small umbra) move towards the light bridge
   (see Fig.~\ref{FigLightbridge}). A high density of UDs is formed by splitting off the light bridge,
   but all in one direction, in which the LB is corrugated and unsharp (disk center, large umbra side),
   while on its other straight and sharp side nearly no UDs leave the LB. 

   \begin{table*}
   \caption{Characteristic values of UD parameters for different UD classes defined in the main text.}
   \label{CharValues}                                    
   \centering                                            
   \begin{tabular}{l l l l l l l l l l}                  
   \hline                                                
   \noalign{\smallskip}
   Class               & Condition       & $N$        & $D^{Mean}$           & $v^{Mean}$            & $I_{Peak}^{Mean}$       & $(I_{Peak}/I_{bg})^{Mean}$ & $T$                    & $L_{BD}$               & $L_{Traj}$             \\
   \noalign{\smallskip}                                                                                                                                                                                                               
                       &                 &            & [km]                 & [m\,s$^{-1}$]         & [$I_{ph}$]              &                            & [s]                    & [km]                   & [km]                   \\
   \noalign{\smallskip}                                                                                                                                                                                                                  
   \hline                                                                                                                                                                                                                                
   \noalign{\smallskip}                                                                                                                                                                                                                  
      all UDs          &                 & \tt{12836} & \tt{229}$\pm$\tt{68} &                       & \tt{O.51}$\pm$\tt{O.O9} & \tt{1.1O}$\pm$\tt{O.O8}    & \tt{~18O}$\pm$\tt{39O} & \tt{~~5O}$\pm$\tt{13O} & \tt{~~7O}$\pm$\tt{2OO} \\
      all UDs          & $T>$\tt{15O}\,s & \tt{~2899} & \tt{272}$\pm$\tt{53} & \tt{42O}$\pm$\tt{19O} & \tt{O.55}$\pm$\tt{O.1O} & \tt{1.17}$\pm$\tt{O.1O}    & \tt{~63O}$\pm$\tt{63O} & \tt{~19O}$\pm$\tt{22O} & \tt{~29O}$\pm$\tt{35O} \\
      peripheral UDs   & $T>$\tt{15O}\,s & \tt{~~621} & \tt{252}$\pm$\tt{44} & \tt{45O}$\pm$\tt{21O} & \tt{O.65}$\pm$\tt{O.O7} & \tt{1.23}$\pm$\tt{O.11}    & \tt{~56O}$\pm$\tt{55O} & \tt{~21O}$\pm$\tt{25O} & \tt{~29O}$\pm$\tt{35O} \\
      central UDs      & $T>$\tt{15O}\,s & \tt{~2278} & \tt{278}$\pm$\tt{53} & \tt{41O}$\pm$\tt{19O} & \tt{O.53}$\pm$\tt{O.O9} & \tt{1.15}$\pm$\tt{O.O9}    & \tt{~65O}$\pm$\tt{65O} & \tt{~18O}$\pm$\tt{21O} & \tt{~3OO}$\pm$\tt{35O} \\
      mobile UDs       & $T>$\tt{15O}\,s & \tt{~~~85} & \tt{287}$\pm$\tt{33} & \tt{68O}$\pm$\tt{14O} & \tt{O.64}$\pm$\tt{O.O8} & \tt{1.29}$\pm$\tt{O.O9}    & \tt{227O}$\pm$\tt{83O} & \tt{1O8O}$\pm$\tt{3OO} & \tt{147O}$\pm$\tt{41O} \\
      stationary UDs   & $T>$\tt{15O}\,s & \tt{~2814} & \tt{272}$\pm$\tt{53} & \tt{41O}$\pm$\tt{19O} & \tt{O.55}$\pm$\tt{O.1O} & \tt{1.16}$\pm$\tt{O.1O}    & \tt{~58O}$\pm$\tt{55O} & \tt{~16O}$\pm$\tt{15O} & \tt{~26O}$\pm$\tt{27O} \\
      chain A UDs      & $T>$\tt{15O}\,s & \tt{~~~48} & \tt{3O1}$\pm$\tt{45} & \tt{48O}$\pm$\tt{19O} & \tt{O.68}$\pm$\tt{O.O6} & \tt{1.29}$\pm$\tt{O.O8}    & \tt{~96O}$\pm$\tt{71O} & \tt{~24O}$\pm$\tt{23O} & \tt{~48O}$\pm$\tt{4OO} \\ 
      light bridge UDs & $T>$\tt{15O}\,s & \tt{~~217} & \tt{254}$\pm$\tt{41} & \tt{43O}$\pm$\tt{19O} & \tt{O.64}$\pm$\tt{O.O8} & \tt{1.23}$\pm$\tt{O.11}    & \tt{~8OO}$\pm$\tt{8OO} & \tt{~22O}$\pm$\tt{24O} & \tt{~37O}$\pm$\tt{43O} \\ 
   \noalign{\smallskip}
   \hline                                                
   \end{tabular}
   \end{table*}

   The mean diameter $D^{Mean}$, mean horizontal velocity $v^{Mean}$, mean peak intensity $I_{Peak}^{Mean}$,
   mean intensity contrast $(I_{Peak}/I_{bg})^{Mean}$, lifetime $T$, birth-death distance $L_{BD}$, trajectory
   length $L_{Traj}$ (see definitions of these quantities in the previous text), and the number of UDs $N$ of
   the types or classes mentioned earlier in this section are summarized in Table~\ref{CharValues}. We averaged
   over all trajectories of a UD class. The standard deviation $\sigma$ is given after each average value.
   As mentioned earlier, we consider only UDs that lived longer than 150~s in all cases in which we calculate
   the mean velocity; only the first line includes all UDs. The last two rows consider only UDs that are
   close to chain A or the light bridge, respectively. According to Table~\ref{CharValues} the difference
   between peripheral UDs and central UDs is not so large (using the simple categorization described above).
   The largest relative difference is in the brightness. For all other parameters the difference is less
   than $1\sigma$. In contrast to that, the difference between mobile UDs and stationary UDs is more
   significant. A relatively large difference of more than $1\sigma$ is found for the mean horizontal
   velocity, for the mean intensity contrast, for the lifetime, and, again for the brightness. On average
   mobile UDs are brighter, they move faster, and they live longer than stationary UDs but they have similar
   sizes. 

   Sometimes one can observe complete chains of successive UDs that are very close to each other (see label
   A in Fig.~\ref{FigBestImage} and the left panel of Fig.~\ref{FigUdChainVersions}). According to
   Table~\ref{CharValues} these UDs are relatively bright and they move along the chain from the endpoints
   of the chain towards its center, where they disappear. Their mean peak intensity is 0.68~$I_{ph}$ which
   is significantly higher than 0.51~$I_{ph}$, the average of all UDs, but is comparable to that of the
   peripheral and mobile UDs. The mean diameter as well as the mean velocity of the UDs within the chain
   is slightly above average. Table~\ref{CharValues} also reveals that UDs that are born close to the
   light bridge show on average a significantly higher brightness and contrast than the mean UD but all
   other properties do not show remarkable differences.

\begin{figure}
   \centering
   \includegraphics[width=\linewidth-1mm]{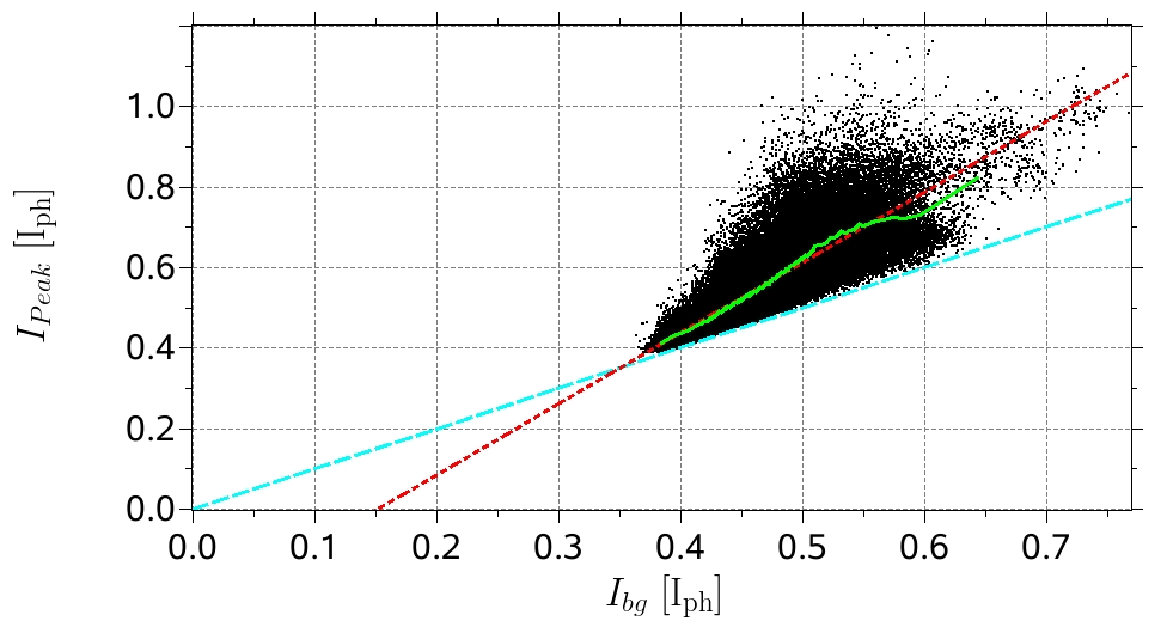}
   \caption{UD peak intensity versus umbral background intensity. The solid green line connects binned values.
   The dashed cyan line displays the theoretical, lower limit of the peak intensities. A linear fit to the data
   results in the dotted red line.}
   \label{FigIPeakVsIbg}
   \end{figure}

   The umbral background is brightest near the penumbra and gets darker towards the center of the umbra.
   Fig.~\ref{FigIPeakVsIbg} shows that the UD peak intensity correlates with the umbral background
   intensity, with the mean ratio $I_{Peak}/I_{bg}$~=~1.2~$\pm$~0.1 (intensity contrast). \citet{Sobotka2005}
   found an intensity contrast of 1.8 for a wavelength of 451~nm and 1.6 for the wavelength 602~nm. Obviously,
   the intensity contrast not only depends on the wavelengths but also on the spatial resolution
   that can be slightly different even if we compare data from the same telescope. Additionally,
   the intensity contrast can also be affected by the subsonic filter and the MFBD image restoration we applied.
   Importantly, however,
   both binning the points in Fig.~\ref{FigIPeakVsIbg} (solid green curve) and a linear regression (dotted red curve)
   indicate that the contrast increases nearly linearly with $I_{bg}$. A scatter plot of the UD intensity
   contrast versus the shortest distance between the UD birth position and the penumbra (not shown) reveals
   that this UD contrast is not constant over the umbra. The closer the UD is born to the penumbra the higher
   its intensity contrast, although the contrast does not drop as rapidly from the umbral boundary as the
   UD brightness does, so that partly, the dependence on distance is due to the dependence on $I_{bg}$
   (Fig.~\ref{FigIPeakVsIbg}). We also found that on average the long-lived UDs have a higher contrast than
   the short-lived ones.

   In a further step we subdivided the umbra into several boxes and determined the probability that a UD
   is born in such a box. The map obtained in this manner shows a uniform distribution of the UD birth
   probability (not plotted). Only for very small box sizes do the dark umbral nuclei (see left panel of
   Fig.~\ref{FigUdOverview}) become visible as locations of reduced UD production.

\begin{figure*}
   \centering
   \includegraphics[width=0.5\linewidth-4mm]{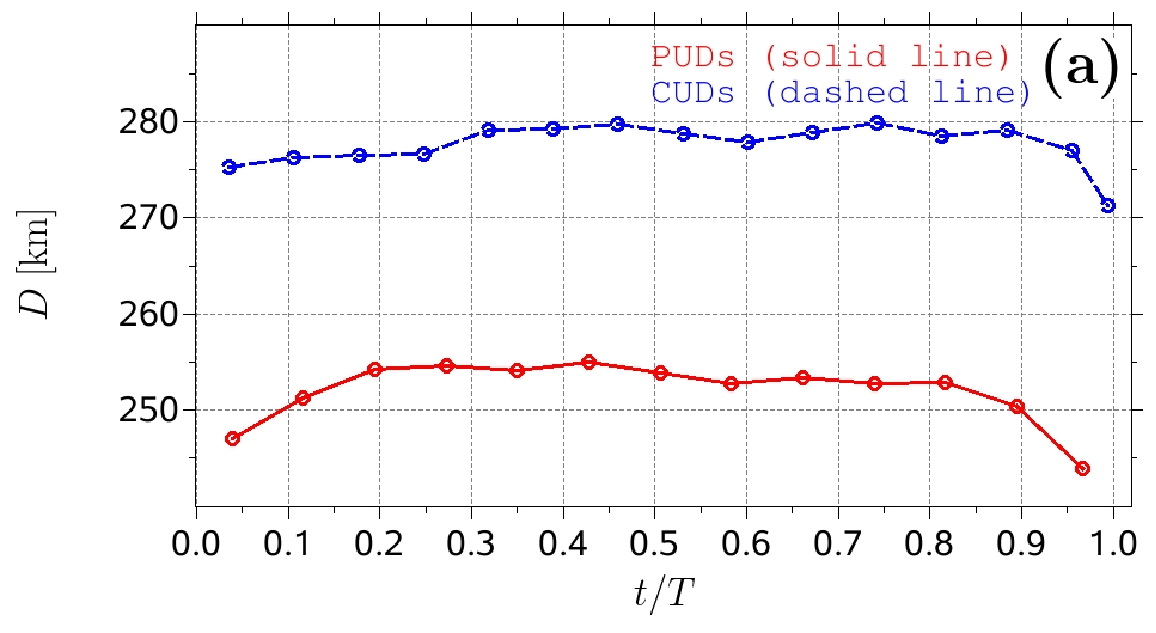}
   \hspace{5mm}
   \includegraphics[width=0.5\linewidth-4mm]{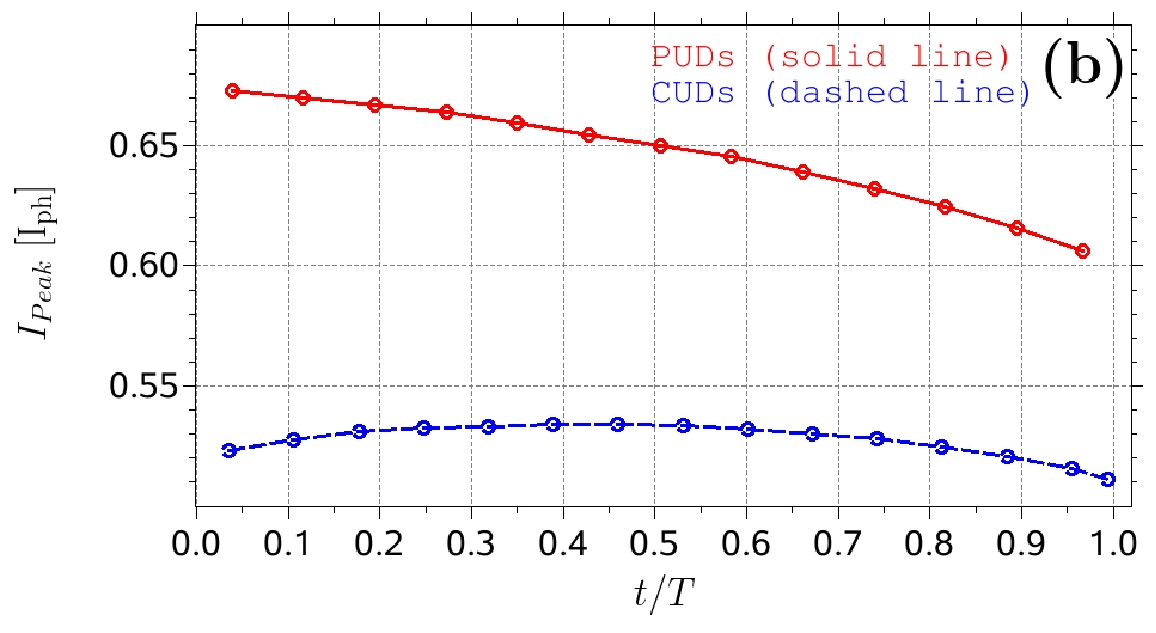}\\[5mm]
   \includegraphics[width=0.5\linewidth-4mm]{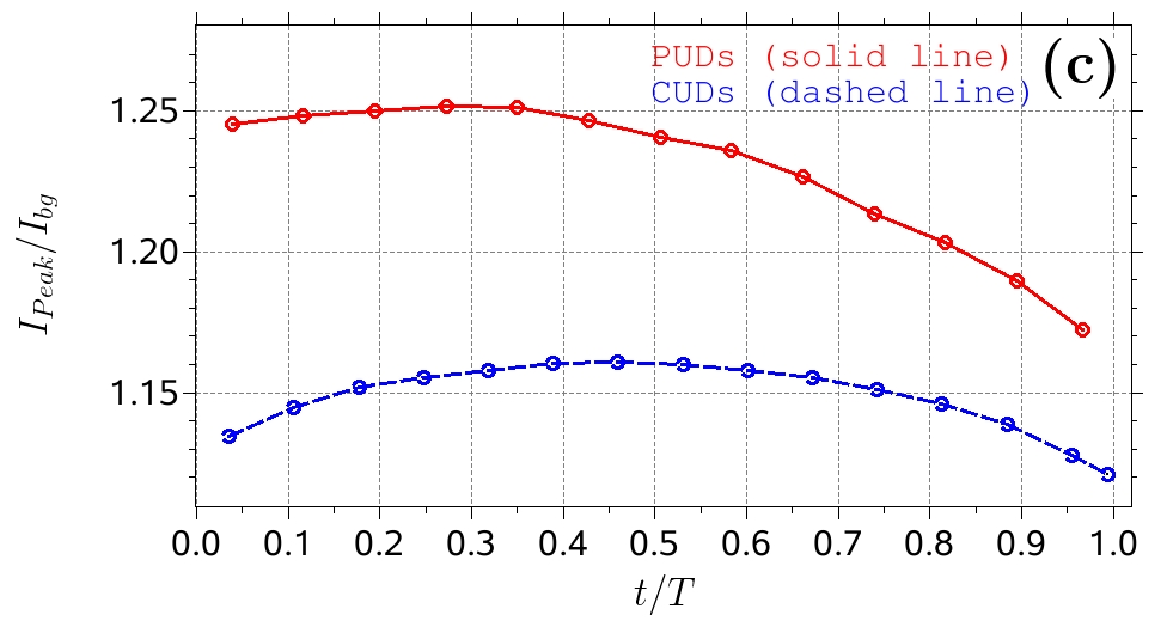}
   \hspace{5mm}
   \includegraphics[width=0.5\linewidth-4mm]{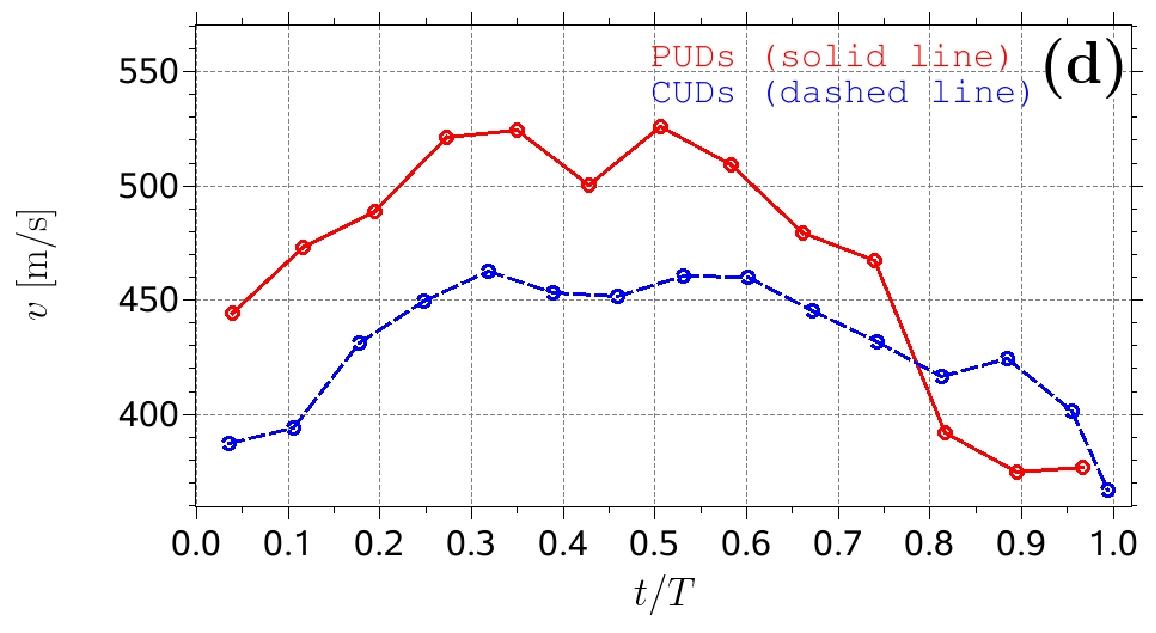}
   \caption{Temporal evolution of the UD diameter (a), UD peak intensity (b), intensity contrast (c), and
   velocity (d) separated for peripheral UDs (solid red line) and central UDs (dashed blue line).
   The UD lifetime is normalized to unity.}
   \label{FigTempEvolution}
   \end{figure*}

   Finally, we are interested in the temporal evolution of the UD properties over their lifetimes. To this
   end we normalize all UD lifetimes to unity and average the temporal evolution of the diameters, peak
   intensities, intensity contrasts ($I_{Peak}/I_{bg}$), and horizontal velocities of the 2899 trajectories
   with $T~>~150~s$. In order to weight all trajectories equally, i.e. independently of their lifetime, we
   up-sample all trajectories to 310 points of time via interpolation. (Our time series contains 310 images,
   so that no trajectory can consist of more than 310 points.) Then we averaged the UD parameters with the
   help of our binning method, which is applied separately for the 621 PUDs and the 2278 CUDs.
   Each bin contains 15000 points for PUDs and 50000 points for CUDs. The results are plotted in
   Fig.~\ref{FigTempEvolution} and show that the mean PUD is smaller, brighter and moves faster than the
   mean CUD (as could already be deduced from Table~\ref{CharValues}). More importantly, there are distinct
   differences in their mean evolution. Whereas both types of UDs share the property that their diameters
   evolve rather moderately over time (the increase after birth and the decrease before death are less than
   10\,\% of the maximum diameter (see panel a), the evolution of their brightness (panel b) and in
   particular of their contrast (panel c) differ considerably. While the mean CUD displays an initial
   gentle brightening followed by an equally gentle darkening, the mean PUD darkens continuously. The small
   magnitude of the change in brightness may be due to the fact that we have averaged over UDs with very
   different absolute intensity. More information may be gleaned from the contrast, i.e. the peak intensity
   divided by the local umbral background intensity, plotted in panel (c). The mean PUD initially remains
   almost constant, exhibiting a slight maximum at around 1/3 of the mean lifetime before dropping
   rapidly over the remaining portion of its life. The contrast of the CUDs displays a much more symmetric
   evolution. The birth velocity of the mean PUD is nearly 50~m\,s$^{-1}$ higher than for the mean CUD,
   while the velocity at death of the two UD types is similar, see panel (d). Both velocity curves show
   an initial increase, followed by a decrease. As in the case of the contrast the velocity profile is
   much more symmetric for the CUDs.

\section{Discussion and conclusions}

   We have analyzed a time series of images of a mature sunspot close to solar disk center.
   Due to the excellent image quality we were able to resolve thousands of UDs. Exhaustive UD analyses
   can be found in earlier papers \citep{Sobotka1997a,Sobotka1997b,Hartkorn2003,Sobotka2005,Sobotka2006},
   but the present article is the first detailed UD study of a long time series of reconstructed images
   with the consistent high resolution of a 1-meter telescope.

   Trajectories, lifetimes, diameters, horizontal velocities, peak intensities, and distances between
   birth and death locations were determined by tracking single UDs over the time series. These
   characteristic values were used to look for reasonable separations into UD classes. In the following
   we summarize the obtained results and compare them with other investigations in the literature:

   \begin{enumerate}
      \item There is hardly any part of the umbra which does not support UDs, but the UD brightnesses
      depend strongly on the location within the umbra, which confirms the previous observation of
      \citet{Sobotka1997b}.
      \item The histogram of lifetimes shows an exponential distribution, i.e. a UD does not have a
      typical lifetime. More than 3/4 of all studied UDs lived less than 150~s and their motion was
      negligible. The exponential distribution is in qualitative agreement with the results obtained by
      \citet{Sobotka1997a,Sobotka1999}. Quantitatively, \citet{Sobotka1997a} obtains a median lifetime
      of 6~min for an umbra of about 6~Mm diameter and a median of 12~min for a 4~Mm pore
      \citep{Sobotka1999}, whereas we find a median value of 0.7~min for a roughly 10~Mm umbra. Note
      that these median values depend strongly on algorithmic constraints as well as on the cadence of
      the time series. For example, the method used by \citet{Sobotka1997a} cannot lead to lifetimes
      shorter than 1.5~min. If we only consider UDs with lifetimes greater than 1.5~min our median
      increases to 4.1~min. Irrespectively of which of these two values we use, our results are
      consistent with the conclusion of \citet{Sobotka1999} that UDs are more stable in a weak magnetic
      field if we assume a direct correlation between umbral diameter and magnetic field strength
      \citep[cf.][]{Kopp1992}. Alternatively, due to the strong dependence of umbral brightness on
      umbral size \citep{Mathew2007} the lifetime may be influenced mainly by the radiative flux or
      umbral temperature. The scatterplot of the mean umbral background intensities versus the UD
      lifetimes (not shown) is consistent with both possible explanations mentioned above: UDs live longer
      in brighter parts of the umbra. Due to the non-linear, monotonically decreasing relation between
      magnetic field strength and background intensity as observed by \citet{Kopp1992} and confirmed
      by \citet{Martinez1993,Solanki1993}, this implies that UDs live longer in regions of weak field.
      \item The histogram of mean diameters exhibits a maximum at 225~km (0.31$^{\prime\prime}$) and
      descends from there towards the diffraction limit, so that we expect the majority of UDs to have
      been spatially resolved. This seems not to be the case in many of the earlier papers because
      there a monotonic decrease was obtained towards higher diameters
      \citep[see][]{Sobotka1997a,Sobotka1999}. \citet{Sobotka2005} also analyzed data obtained with
      the 1-meter SST. These data lead to a histogram that is qualitatively similar to ours. The mean
      diameter of 175~km (0.24$^{\prime\prime}$), as well as the average filling factor of 9\,\% is,
      however, noticeable smaller. This difference can be explained by the use of a different method
      to determine the UD boundary. An increase of our brightness threshold to determine the UD
      boundary leads to smaller diameters and to lower filling factors. \citet{Hamedivafa2008} used
      an improved method of image segmentation and also found a mean diameter of 230~km and a similar
      shaped histogram.
      \item The mean horizontal velocity of those of our UDs that live longer than 150~s is 420~m\,s$^{-1}$
      which is significantly higher than the 210~m\,s$^{-1}$ reported by \citet{Molowny1994} and higher than
      the 320~m\,s$^{-1}$ found by \citet{Sobotka1999}. In both studies the mean horizontal velocity
      was calculated by means of least-squares linear fits of the x and y coordinates of all trajectory
      points, which leads to an underestimation of velocity in case of curved trajectories. Our
      histogram of horizontal velocities shows a maximum at 350~m\,s$^{-1}$, whereas some UDs can reach
      velocities above 1~km\,s$^{-1}$. This is in qualitative agreement with the histograms of
      \citet{Kitai1986}, \citet{Molowny1994}, and \citet{Hamedivafa2008} but disagrees with the results of
      \citet{Sobotka1997b,Sobotka1999} whose histograms do not exhibit a maximum; they peak at zero
      velocity and show a monotonic decrease toward higher velocities of up to 1~km\,s$^{-1}$.
      The majority of our UDs moves irregularly around the birth position. However, there are some
      mobile UDs that travel over long distances within their lifetime. Almost all mobile UDs emerge
      close to the umbral border, i.e. near the penumbra, they are brighter than the average and their
      horizontal motion is preferentially directed towards the center of the umbra. The mean velocity
      of our mobile UDs is 680~m\,s$^{-1}$, which is in good agreement with the recent observation
      of \citet{Katsukawa2007} who found a mean velocity of 700~m\,s$^{-1}$.
      \item The relation between mean UD size and lifetime is non-linear. On average, the size of UDs
      increases with lifetime, which was also found by \citet{Sobotka1997a}, but in contrast to their
      work we find a narrow size distribution of around 290~km for long-lived UDs.
      \item UDs that were born close to the penumbra show a significantly higher contrast than
      the UDs of the umbral interior. The mean UD intensity contrast $I_{Peak}/I_{bg}$ is 1.2 which is
      smaller than the value of 1.6 reported by \citet{Sobotka2005}. This may partly be due to the
      longer wavelength of our observation. Additionally, our statistical ensemble contains many more UDs.
      In particular we took many UDs with low contrast into account, made possible by the multilevel
      tracking technique \citep{Bovelet2001} we employed to identify UDs. Consequently, we believe that
      the lower UD contrast we find is not due to a lower resolution or higher stray light, but rather
      to differences in identification of UDs and in particular the difference in wavelength. We stress
      that the UD contrast itself depends on the background intensity; the higher the intensity the
      stronger the contrast. Also it cannot be ruled out that there could be systematically
      different contrasts between different sunspots due to intrinsicly different physical properties of
      the spots.
      \item Whereas the temporal variation of the UD diameter is qualitatively similar for UDs formed
      close to the penumbra (PUDs) and those formed in the body of the umbra (CUDs), their intensity
      contrast and horizontal velocity display contrasting evolutions. The mean PUD shows a continuous
      darkening which is in agreement with the results of \citet{Kitai2007} for a single typical PUD.
      The typical CUD of \citet{Kitai2007} is found to increase in brightness linearly and then to darken
      linearly with time, whereas our results for the mean CUD show a non-linear increase in brightness
      until nearby half of the lifetime followed by an again non-linear decrease. The clear difference
      between PUDs and CUDs in the behavior of their contrast and mean velocity may be a result of the
      different origin of the two types of features \citep{Kitai2007}. We confirm from visual inspection
      of a subset of PUDs that PUDs are formed when penumbral grains cross the umbral boundary.
   \end{enumerate}

   A comparison of the results with the simulations of \citet{Schuessler2006} shows a better agreement
   with CUDs, than PUDs. For example, the simulated UDs display a gradual increase in contrast
   followed by a gradual decrease, just as CUDs. They also display little proper motion. This qualitative
   agreement further strengthens the interpretation of UDs as localized columns of overturning convection
   proposed by \citet{Schuessler2006}. Hinode data had earlier suggested the presence of dark lanes in
   large UDs \citep{Bharti2007} and revealed a decrease in the magnetic field strength with depth,
   as well as an upflow associated with a temperature enhancement \citep{Riethmueller2008}, in good
   qualitative agreement with the simulations. A detailed analysis of the simulations similar to the one
   carried out here would allow a more quantitative comparison.

   The PUDs have significantly different evolution histories than the simulated features. They start
   at a higher speed \citep{Kitai1986} and in particular display their maximum brightness right after the
   beginning of their life \citep[cf.][]{Kitai2007}. This, combined with the fact that they are born very close
   to the penumbra, or actually by breaking away from the penumbra \citep{Thomas2004}, and move radially
   towards the umbral center \citep{Kitai1986} supports that these are two distinct types of UDs based on
   their origin and evolution, although their physical structure is relatively similar \citep{Riethmueller2008}.



\begin{thebibliography}{}

   \bibitem[Adjabshirzadeh \& Koutchmy(1983)]{Adjabshirzadeh1983} Adjabshirzadeh, A., \& Koutchmy, S. 1983,
      A\&A, 121, 1

   \bibitem[Barrodale et al.(1993)]{Barrodale1993} Barrodale, I., Skea, D., Berkley, M., Kuwahara, R., \& Poeckert, R. 1993,
      Pattern Recognition, 26, 375

   \bibitem[Berdyugina et al.(2003)]{Berdyugina2003} Berdyugina, S. V., Solanki, S. K., \& Frutiger, C. 2003,
      A\&A, 412, 513

   \bibitem[Berger \& Berdyugina(2003)]{Berger2003} Berger, T. E., \& Berdyugina, S. V. 2003,
      ApJ, 589, L117

   \bibitem[Bharti et al.(2007)]{Bharti2007} Bharti, L., Joshi, C., \& Jaaffrey, S. N. A. 2007,
      ApJ, 669, L57

   \bibitem[Bovelet \& Wiehr(2001)]{Bovelet2001} Bovelet, B., \& Wiehr, E. 2001,
      Solar Phys., 201, 13  

   \bibitem[Choudhury(1986)]{Choudhury1986} Choudhury, A. R. 1986,
      ApJ, 302, 809

   \bibitem[Grossmann-Doerth et al.(1986)]{Grossmann1986} Grossmann-Doerth, U., Schmidt, W., \& Schr\"oter, E. H. 1986,
      A\&A, 156, 347

   \bibitem[Hamedivafa(2008)]{Hamedivafa2008} Hamedivafa, H. 2008,
      Solar Phys., in press  

   \bibitem[Hartkorn \& Rimmele(2003)]{Hartkorn2003} Hartkorn, K., \& Rimmele, T. 2003,
      in Current Theoretical Models and Future High-Resolution Solar Observations: Preparing for ATST,
      ed.\ A. A. Pevtsov, \& H. Uitenbroek,
      ASP Conf. Ser., Vol. 286, p. 281  

   \bibitem[Hirzberger et al.(2002)]{Hirzberger2002} Hirzberger, J., Bonet, J. A., Sobotka, M., V\'azquez, M., \& Hanslmeier, A. 2002,
      A\&A, 383, 275

   \bibitem[Katsukawa et al.(2007)]{Katsukawa2007} Katsukawa, Y., Yokoyama, T., Berger, T. E., et al. 2007,
      PASJ, 59, S577

   \bibitem[Kitai(1986)]{Kitai1986} Kitai, R. 1986,
      Solar Phys., 104, 287  

   \bibitem[Kitai et al.(2007)]{Kitai2007} Kitai, R., Watanabe, H., Nakamura, T., et al. 2007,
      PASJ, 59, S585

   \bibitem[Kopp \& Rabin(1992)]{Kopp1992} Kopp, G., \& Rabin, D. 1992,
      Solar Phys., 141, 253  

   \bibitem[Langhans et al.(2007)]{Langhans2007} Langhans, K., Scharmer, G. B., Kiselman, D., \& L\"ofdahl, M. G. 2007,
      A\&A, 464, 763

   \bibitem[Lites et al.(2004)]{Lites2004} Lites, B. W., Scharmer, G. B., Berger, T. E., \& Title, A. M. 2004,
      Solar Phys., 221, 65

   \bibitem[L\"ofdahl(2003)]{Loefdahl2003} L\"ofdahl, M. G. 2003,
      in Image Reconstruction from Incomplete Data II,
      ed.\ P. J. Bones, M. A. Fiddy, \& R. P. Millane,
      Proc. of the SPIE, Vol. 4792, p. 146

   \bibitem[Mart\'inez \& V\'azquez(1993)]{Martinez1993} Mart\'inez Pillet, V., \& V\'azquez, M. 1993,
      A\&A, 270, 494

   \bibitem[Mathew et al.(2007)]{Mathew2007} Mathew, S. K., Mart\'inez Pillet, V., Solanki, S. K., \& Krivova, N. A. 2007,
      A\&A, 465, 291

   \bibitem[Molowny-Horas(1994)]{Molowny1994} Molowny-Horas, R. 1994,
      Solar Phys., 154, 29  

   \bibitem[November(1988)]{November1988} November, L. J., \& Simon, G. W. 1988,
      ApJ, 333, 427

   \bibitem[Parker(1979)]{Parker1979} Parker, E. N. 1979,
      ApJ, 234, 333

   \bibitem[Riethm\"uller et al.(2008)]{Riethmueller2008} Riethm\"uller, T. L., Solanki, S. K., \& Lagg, A. 2008,
      ApJ, 678, L157

   \bibitem[Rimmele(2008)]{Rimmele2008} Rimmele, T. 2008,
      ApJ, 672, 684

   \bibitem[Scharmer et al.(2002)]{Scharmer2002} Scharmer, G. B., Gudiksen, B. V., Kiselman, D., L\"ofdahl, M. G.,\& Rouppe van der Voort, L. 2002,
      Nature, 420, 151

   \bibitem[Scharmer et al.(2003a)]{Scharmer2003a} Scharmer, G. B., Bjelksjo, K., Korhonen, T. K., Lindberg, B.,\& Petterson, B. 2003a,
      in Innovative Telescopes and Instrumentation for Solar Astrophysics,
      ed.\ S. L. Keil, \& S. V. Avakyan,
      Proc. of the SPIE, 4853, 341

   \bibitem[Scharmer et al.(2003b)]{Scharmer2003b} Scharmer, G. B., Dettori, P. M., L\"ofdahl, M. G., \& Shand, M. 2003b,
      in Innovative Telescopes and Instrumentation for Solar Astrophysics,
      ed.\ Stephen L. Keil,\& Sergey V. Avakyan,
      Proc. of the SPIE, 4853, 370

   \bibitem[Scharmer et al.(2007)]{Scharmer2007} Scharmer, G. B., Langhans, K., Kiselman, D.,\& L\"ofdahl, M. G. 2007,
      in New Solar Physics with the Solar-B Mission,
      ed.\ K. Shibata, S. Nagata, \& T. Sakurai,
      ASP Conf. Ser., Vol. 369, p. 71

   \bibitem[Sch\"ussler \& V\"ogler(2006)]{Schuessler2006} Sch\"ussler, M., \& V\"ogler, M. 2006,
      ApJ, 641, L73

   \bibitem[Sobotka(2006)]{Sobotka2006} Sobotka, M. 2006,
      Dissertation for Doctor Scientiarum, Acad. Sci. Czech Republic

   \bibitem[Sobotka et al.(1997a)]{Sobotka1997a} Sobotka, M., Brandt, P. N., \& Simon, G. W. 1997a,
      A\&A, 328, 682

   \bibitem[Sobotka et al.(1997b)]{Sobotka1997b} Sobotka, M., Brandt, P. N., \& Simon, G. W. 1997b,
      A\&A, 328, 689

   \bibitem[Sobotka \& Hanslmeier(2005)]{Sobotka2005} Sobotka, M., \& Hanslmeier, A. 2005,
      A\&A, 442, 323

   \bibitem[Sobotka et al.(1999)]{Sobotka1999} Sobotka, M., V\'azquez, M., Bonet, J. A., Hanslmeier, A., \& Hirzberger, J. 1999,
      A\&A, 511, 436

   \bibitem[Solanki(2003)]{Solanki2003} Solanki, S. K. 2003,
      A\&A Rev., 11, 153

   \bibitem[Solanki et al.(1993)]{Solanki1993} Solanki, S. K., Walther, U., \& Livingston, W. 1993,
      A\&A, 277, 639

   \bibitem[Thomas \& Weiss(2004)]{Thomas2004} Thomas, J. H., \& Weiss, N. O. 2004,
      ARA\&A, 42, 517

   \bibitem[Title et al.(1989)]{Title1989} Title, A. M., Tarbell, T. D., Topka K. P., Ferguson, S. H., \& Shine, R. A. 1989,
      ApJ, 336, 475

   \bibitem[Tritschler \& Schmidt(2002)]{Tritschler2002} Tritschler, A., \& Schmidt, W. 2002,
      A\&A, 388, 1048

   \bibitem[Weiss et al.(1990)]{Weiss1990} Weiss, N. O., Brownjohn, D. P., Hurlburt, N. E., \& Proctor, M. R. E. 1990,
      MNRAS, 245, 434

   \bibitem[Zakharov et al.(2008)]{Zakharov2008} Zakharov, V., Hirzberger, J., Riethm\"uller, T. L., Solanki, S. K., \& Kobel, P. 2008,
      A\&A, 488, L17

\end{thebibliography}
\end{document}